\documentclass[journal,finalcls,twocolumn,10pt,twoside]{IEEEtran}
\normalsize

\usepackage{cite}

\ifCLASSINFOpdf
   \usepackage[pdftex]{graphicx}
\else
\fi

\usepackage[cmex10]{amsmath}
\usepackage{array}

\usepackage{mdwmath}
\usepackage{mdwtab}

\usepackage{eqparbox}

\usepackage[tight,footnotesize]{subfigure}

\usepackage{fixltx2e}

\usepackage{enumerate}
\usepackage{epstopdf} 
\usepackage{multirow}
\usepackage{amsmath,amsthm,amsfonts,amssymb}
\usepackage{diagbox}
\usepackage{xcolor}
\usepackage{hhline}
\usepackage{dblfloatfix} 

\newcounter{MYtempeqncnt}

\begin{document}
%
\title{Inter-Technology Coexistence in a \\Spectrum Commons: A Case Study of \\Wi-Fi and LTE in the 5 GHz Unlicensed Band}
%
%
%

\author{Andra~M.~Voicu, Ljiljana Simi\'c and Marina Petrova
\thanks{A.~M.~Voicu, L.~Simi\'c, and M.~Petrova are with the Institute for Networked Systems, RWTH Aachen University (e-mail: avo@inets.rwth-aachen.de; lsi@inets.rwth-aachen.de; mpe@inets.rwth-aachen.de).}}

%
%

\markboth{}
{}
%



\maketitle

\begin{abstract}

Spectrum sharing mechanisms need to be carefully designed
to enable \mbox{inter-technology} coexistence in the unlicensed bands, as these bands are an instance of a \emph{spectrum commons} where highly heterogeneous technologies and deployments must coexist. 
Unlike in licensed bands, where multiple technologies could coexist only in a primary-secondary DSA mode, a spectrum commons offers competition opportunities between multiple dominant technologies, such as \mbox{Wi-Fi} and the recently proposed LTE in the 5~GHz unlicensed band. 
In this paper we systematically study the performance of different spectrum sharing schemes for \mbox{inter-technology} coexistence in a spectrum commons. 
Our contributions are threefold. 
Firstly, we propose a general framework for transparent comparative analysis of spectrum sharing mechanisms in time and frequency, by studying the effect of key constituent parameters.
Secondly, we propose a novel throughput and interference model for \mbox{inter-technology} coexistence,
integrating \mbox{per-device} specifics of different distributed MAC sharing mechanisms in a unified \mbox{network-level} perspective.
Finally, we present a case study of IEEE 802.11n \mbox{Wi-Fi} and LTE in the 5~GHz unlicensed band, in order to obtain generalizable insight into coexistence in a spectrum commons.
Our extensive Monte Carlo simulation results show that LTE/\mbox{Wi-Fi} coexistence in the 5~GHz band can be ensured simply through channel selection schemes, such that \mbox{time-sharing} MAC mechanisms are irrelevant.
We also show that, in the general \mbox{co-channel} case, the coexistence performance of MAC sharing mechanisms strongly depends on the \emph{interference coupling} in the network, predominantly determined by building shielding. 
We thus identify two regimes: (i)~low interference coupling, e.g. residential indoor scenarios, where duty cycle mechanisms outperform \mbox{sensing-based} \mbox{listen-before-talk} (LBT) mechanisms; and (ii) high interference coupling, e.g. \mbox{open-plan} indoor or outdoor hotspot scenarios, where LBT outperforms duty cycle mechanisms.

\end{abstract}


%
\IEEEpeerreviewmaketitle

\section{Introduction}
%
%
%
%

With the densification of heterogeneous wireless-capable devices and the rapid and continuous increase in data traffic volumes in wireless networks~\cite{Cisco2016}, spectrum sharing techniques are essential for mitigating mutual interference between \mbox{co-located}, \mbox{co-channel} wireless devices, thereby enabling concurrent operation of multiple devices.
It follows that, in practice, the technical design of spectrum sharing techniques for a given technology depends on three major aspects: (i)~the technologies implemented by the other devices, where interference is to be managed either between devices of the same technology (i.e. \mbox{intra-technology} coexistence), or of different technologies (i.e. \mbox{inter-technology} coexistence); (ii)~the management of the devices, where interference may be managed with various levels of coordination (i.e. intra- and \mbox{inter-operator} coexistence), or in a fully distributed manner (for individually deployed devices); and (iii)~the management of the spectrum, spanning a continuum of access models, from exclusive use of spectrum (i.e. exclusive spectrum access rights for a single operator/technology) to a spectrum commons (i.e. equal spectrum access rights for all users/operators/technologies)~\cite{Peha2005}.   

From a regulatory and economical perspective, spectrum can be broadly classified into licensed 
and unlicensed 
bands. 
Licensed bands are typically associated with exclusive spectrum access, such as for cellular networks that traditionally implement \mbox{intra-technology}, \mbox{intra-operator} coordinated spectrum sharing techniques. This case is thus less challenging for designing spectrum sharing mechanisms. 
Although not yet widely implemented, dynamic spectrum access (DSA)~\cite{Liang2011}, can enable inter-technology coexistence in licensed bands, where one primary technology is dominant and the other secondary technologies are required to give access priority to the primary. The design of spectrum sharing mechanisms in such a case is limited by the given priority constraints of the primary license holder. 
The unlicensed bands are a prominent example of a \emph{spectrum commons}, since they are in principle open for any type of device management and technology\footnote{As long as basic regulatory limitations, e.g. spectrum transmission masks, are complied with.}. Importantly, the broader concept of spectrum commons refers only to equal spectrum access rights, disregarding economical aspects, so licensed spectrum could also hypothetically be a spectrum commons for more technologies/operators with equal access rights on a cost basis~\cite{Peha2005}.      

To facilitate coexistence of heterogeneous devices in dense deployments, the design of novel spectrum sharing techniques for the unlicensed bands (i.e. MAC layer mechanisms, channel allocation, etc.) has to consider other legacy techniques operating concurrently. Also, unlike in the licensed bands, in the unlicensed bands any technology has the same access priority rights, such that more than one distinct technology can be dominant.
The \mbox{sub-6}~GHz unlicensed bands have consistently been an innovation driver for technology development, as 
they are ideally suited to sporadic transmissions in, e.g. emerging M2M and IoT applications~\cite{Andreev2015}, while accommodating loaded legacy and emerging small cell networks. However, with the continuous proliferation of devices and technologies in these bands, it is essential to consider the effectiveness of spectrum sharing mechanisms, in order to maximize overall utilization efficiency, while exploiting the low entry barrier for new technologies. 
Moreover, the high level of heterogeneity in these bands offers the opportunity to generalize \mbox{inter-technology} coexistence models applicable to future spectrum commons, e.g. the recently opened 3.5~GHz band for wireless broadband applications~\cite{FederalCommunicationsCommission2015}.  

IEEE 802.11 \mbox{Wi-Fi} is currently the only dominant technology in the \mbox{sub-6}~GHz unlicensed bands, in terms of data traffic volumes and the number of users. Although Bluetooth and IEEE 802.15.4 coexist with \mbox{Wi-Fi} in the 2.4~GHz band, these technologies are mainly used for \mbox{short-range} sporadic communications, and therefore \mbox{inter-technology} coexistence is easily ensured. 
However, the recently proposed LTE in the unlicensed bands~\cite{Qualcomm2015a} would be a \emph{second} dominant technology coexisting with \mbox{Wi-Fi} in the 5~GHz band. 
LTE would thus have to share the spectrum with \mbox{Wi-Fi} (i.e. \mbox{inter-technology} spectrum sharing) and other LTE systems (i.e. \mbox{inter-operator}, \mbox{intra-technology} spectrum sharing), despite originally having been designed for exclusive use of licensed spectrum. 
The main coexistence issue arises from the difference in the design of the MAC  layer mechanisms that \mbox{Wi-Fi} and LTE implement. The \mbox{Wi-Fi} MAC is based on CSMA/CA, which is a \mbox{listen-before-talk} (LBT)  mechanism where a device senses the medium and allows other devices to finish their transmission before starting its own transmission. Unlike \mbox{Wi-Fi}, LTE transmits almost continuously without checking whether the medium is occupied, as it traditionally exclusively operates on its own portion of spectrum. Such a behaviour in the unlicensed band would completely prevent \mbox{Wi-Fi} transmissions and would cause strong interference to other LTE networks operating in the same band~\cite{Babaei2014}.
Consequently, new spectrum sharing techniques for LTE must be designed and evaluated, in order to ensure harmonious inter- and intra-technology coexistence. 

In this paper we systematically study the performance of different spectrum sharing mechanisms in a spectrum commons, with two dominant technologies. 
We consider two types of access point (AP) populations, i.e. \emph{legacy APs} and \emph{new entrant APs}, and we conduct a detailed \mbox{system-level} coexistence study for multiple scenarios and a wide range of realistic AP densities. 
We consider LTE/\mbox{Wi-Fi} coexistence as a case study and we assume legacy \mbox{Wi-Fi} APs with an LBT spectrum sharing mechanism, whereas for the new entrant LTE APs we consider various candidate spectrum sharing mechanisms: LBT with different fixed carrier sense (CS) thresholds, different variations of fixed and adaptive duty cycle, and different channel selection schemes. 
Our contributions are threefold.

Firstly, we propose a general framework  with the goal of enabling comparative assessment of spectrum sharing mechanisms when several networks of heterogeneous technologies coexist in a spectrum commons, by identifying the key constituent parameters of spectrum sharing schemes and investigating their individual effect on the coexistence performance.
Some early studies analyzed coexistence between \mbox{Wi-Fi} and Bluetooth~\cite{Lansford2001}, or \mbox{Wi-Fi} and IEEE 802.15.4~\cite{SikoraOttawa2005} in the 2.4 GHz unlicensed band, but their focus was on specific technologies where only one is dominant. Prior studies of \mbox{Wi-Fi} coexistence with different variants of LTE in the unlicensed band, e.g.~\cite{3GPP2015},~\cite{Forum2015}, typically consider a single MAC approach for LTE with dissimilar (and often incompletely specified) system assumptions and scenarios, so that these studies are neither directly  comparable nor readily reproducible. By contrast, we demonstrate a unified and generalizable framework for systematically exploring the design space of spectrum sharing mechanisms and thus transparently evaluating \mbox{inter-technology} coexistence in a spectrum commons.

Secondly, we propose a novel throughput and interference model for heterogeneous technology coexistence in a spectrum commons, detailed enough to capture the key parameters of several MAC sharing mechanisms, while abstract enough to enable meaningful comparison between the MAC mechanisms. 
Earlier throughput and interference models in the literature~\cite{Bianchi2000, H.Q.Nguyen2007, VoicuLondon2015} focus on only one specific technology in simplified scenarios, or on very high-level models that offer only an approximate \mbox{network-level} estimate. 
By contrast, our model is the first one to incorporate the \mbox{per-device} specifics of multiple potential coexisting technologies in a unified \mbox{network-level} perspective, thereby extending and integrating prior approaches.

Thirdly, we present a coexistence case study of \mbox{Wi-Fi} and LTE in the 5~GHz unlicensed band, inspired by its relevance to the contemporary industry and academic context.  Importantly, our discussion of the case study results is generalizable and also gives insight into other possible coexistence cases in a spectrum commons.      
Several spectrum sharing mechanisms have been proposed and initially studied for LTE in the unlicensed bands~\cite{Zhang2015, Nhtilae2013, Almeida2013, Jeon2014, Forum2015, Qualcomm2015, Alliance2015, Alliance2016, Zhang2015a, RupasingheNewOrleans2015, Al-Dulaimi2015, 2015, 3GPP2015, BhorkarNewOrleans2015, JiaLondon2015, VoicuLondon2015, ChenGlasgowMay2015, JeonLondon2015, RatasukAustin2014, Song2016, TaoHongKong2015, LiHongKong2015, Bhorkar2014, SagariLondon2015, SallentLondon2015, Abinader2014, XiaLondon2015}, but most of the work has focused on either optimizing one mechanism for particular network conditions, or analysing very few variations of the same mechanism in simplistic scenarios. Extensive system-level studies analysing multiple fundamental spectrum sharing mechanisms under the same framework, with comparable and generalizable results, are missing from the literature. 
We show in this paper that LTE/\mbox{Wi-Fi} coexistence can be easily ensured simply through channel selection schemes, such that \mbox{time-sharing} MAC mechanisms are irrelevant, given the high number of available channels in the 5~GHz band. 
Importantly, our analysis shows that, in the general \mbox{co-channel} case, the coexistence performance of the MAC strongly depends on the interference coupling, largely determined by building shielding, resulting in two regimes: (i)~low interference coupling, e.g. residential indoor scenarios, where distributed duty cycle mechanisms outperform \mbox{sensing-based} LBT approaches; and (ii) high interference coupling, e.g. \mbox{open-plan} indoor or outdoor hotspot scenarios, where LBT outperforms duty cycle mechanisms. 
We also show that the performance of LBT is close to the performance of perfectly coordinated adaptive duty cycle MAC mechanisms, suggesting that distributed MAC schemes remain more attractive in practice.

The remainder of this paper is organized as follows. Section~\ref{section: related work} summarizes related work in the literature. Section~\ref{section: mechanisms} presents the proposed coexistence evaluation framework. Section~\ref{section: interf} presents our novel throughput and interference model. Section~\ref{section: results} presents and discusses our case study results, and Section~\ref{section: conclusion} concludes the paper.

\section{Related Work}
\label{section: related work}

Previous work has addressed legacy \mbox{inter-technology} coexistence studies in the unlicensed bands~\cite{Lansford2001, SikoraOttawa2005}, throughput and interference models for a spectrum commons~\cite{Bianchi2000, H.Q.Nguyen2007, VoicuLondon2015}, and LTE/\mbox{Wi-Fi} coexistence in the unlicensed bands~\cite{Zhang2015, Nhtilae2013, Almeida2013, Jeon2014, Forum2015, Qualcomm2015, Alliance2015, Alliance2016, Zhang2015a, RupasingheNewOrleans2015, Al-Dulaimi2015, 2015, 3GPP2015, BhorkarNewOrleans2015, JiaLondon2015, VoicuLondon2015, ChenGlasgowMay2015, JeonLondon2015, RatasukAustin2014, Song2016, TaoHongKong2015, LiHongKong2015, Bhorkar2014, SagariLondon2015, SallentLondon2015, Abinader2014, XiaLondon2015}.

\subsubsection*{Inter-technology coexistence studies in the unlicensed bands}
Several studies in the literature have analysed coexistence between \mbox{Wi-Fi} and other legacy technologies operating in the unlicensed bands, such as Bluetooth~\cite{Lansford2001} and ZigBee~\cite{SikoraOttawa2005}. However, these specifically focus on coexistence with existing standardized technologies and the case where \mbox{Wi-Fi} dominates in terms of traffic volumes, coverage range, and deployment scale. By contrast, we take a more generic approach to explore the coexistence design space and systematically study multiple candidate dominant technologies coexisting in a spectrum commons, in order to determine the key parameters that would improve the coexistence performance. 

\begin{table*}[t!]
\begin{center}
\caption{Classification of spectrum sharing mechanisms for LTE in the unlicensed bands \label{table_2} }
\begin{tabular}{|l|l|l|l|l|c|}
\cline{6-6}
 \multicolumn{5}{c|}{} & \textbf{Examples}\\
\hline 
\multirow{31}{*}{\rotatebox[origin=c]{90}{\textbf{Spectrum sharing mechanisms}}} 
 & \multirow{21}{*}{\rotatebox[origin=c]{90}{\textbf{MAC}}}
 & \multirow{10}{*}{\textbf{Duty cycle}} & \multicolumn{2}{l|}{\emph{fixed}}  & 
                   \begin{tabular}{p{7.5cm}l}  20\%-80\% of LTE subframes~\cite{Zhang2015, Nhtilae2013, Almeida2013};
                                               50\% synchronous/asynchronous LTE subframes~\cite{Jeon2014}
                   \end{tabular}
 \\
 \cline{4-6}
 & & & \multicolumn{2}{l|}{\emph{adaptive}}  & 
                   \begin{tabular}{p{7.5cm}l} $\bullet$ cycle range 20-100~ms~\cite{Forum2015} \\
                                            
                                            $\bullet$ CSAT: 80, 160, 640~ms cycle range, max ON duration 100~ms, subframe puncturing 2/20~ms~\cite{Qualcomm2015}; 80 and 400~ms duty cycle period, ON duration 40~ms, subframe puncturing 2/20 or 1/40~ms~\cite{Alliance2015}; ON duration 4-20~ms~\cite{Alliance2016}; \\
                                            $\bullet$ Q-learning with 2~ms granularity and 20~ms period~\cite{RupasingheNewOrleans2015}  \\
                                            $\bullet$ coordinated/uncoordinated duty cycle per subframe~\cite{Al-Dulaimi2015}
                   \end{tabular}
 \\
\hhline{|~|~|====}
 & & \multirow{12}{*}{\textbf{LBT}} & \multirow{8}{*}{\emph{backoff type}} & \emph{no backoff} & 
                   \begin{tabular}{p{7.5cm}l}  ETSI FBE~\cite{2015, 3GPP2015}; sense 34~$\mu$s~\cite{BhorkarNewOrleans2015}; sense 2 symbols or 1 subframe~\cite{JiaLondon2015}; ideal LBT~\cite{VoicuLondon2015} 
                   \end{tabular}
 \\
 		\cline{5-6}
 			& & & & \begin{tabular}{l} \emph{fixed} \\ \emph{contention} \\ \emph{window} \\ \emph{(CW)} \end{tabular} &
 			       \begin{tabular}{p{7.5cm}l}  ETSI LBE option B with CW 4 to 32 and backoff slot of at least 20~$\mu$s~\cite{2015, 3GPP2015}; CW=2, 4~\cite{ChenGlasgowMay2015}; CW=32, 128~\cite{Jeon2014, JeonLondon2015}; CW=2 subframes~\cite{RatasukAustin2014}; CW fixed, but optimized for total throughput maximization~\cite{Song2016}
                   \end{tabular} 
 \\
 		\cline{5-6}
 			& & & & \emph{adaptive CW} & 
 			       \begin{tabular}{p{7.5cm}l}  ETSI LBE option A or 802.11 with binary exponential random backoff~\cite{2015, 3GPP2015}; adaptive CW for QoS fairness~\cite{TaoHongKong2015} 
                  \end{tabular} 
 \\
	\cline{4-6} 
	&  & & \multirow{2}{*}{\emph{CS threshold}} & \emph{fixed} & 
	               \begin{tabular}{p{7.5cm}l}  -60~dBm for 20~MHz channels~\cite{2015}; -62, -68, -72, -77, \mbox{-82}~dBm~\cite{3GPP2015}; -52, -62, -72, -82, -92~dBm~\cite{JeonLondon2015}
                  \end{tabular}
\\
		\cline{5-6}
			& & & & \emph{adaptive} & 
			       \begin{tabular}{p{7.5cm}l}  -80 to -30~dBm to guarantee fair coexistence with \mbox{Wi-Fi} and exploit frequency reuse for LAA~\cite{LiHongKong2015}
                   \end{tabular}
\\
\hhline{|~|=====}
 & \multicolumn{2}{c|}{\multirow{6}{*}{\textbf{Channel selection}}} & \multicolumn{2}{l|}{\emph{random}} & 
                   \begin{tabular}{p{7.5cm}l} 12 channels~\cite{Bhorkar2014}; 3 channels~\cite{SagariLondon2015}; 11 outdoor and 19 indoor channels~\cite{VoicuLondon2015}
                   \end{tabular}
 \\
 \cline{4-6}
 & \multicolumn{2}{c|}{} & \multicolumn{2}{l|}{\emph{distributed}}  & 
                   \begin{tabular}{p{7.5cm}l} avoid other transmissions~\cite{Forum2015, Alliance2016, 3GPP2015}; least interfered at AP or user~\cite{Bhorkar2014}; CSAT~\cite{Qualcomm2015}; avoid Wi-Fi only~\cite{VoicuLondon2015}; \mbox{Q-learning}~\cite{SallentLondon2015}
                   \end{tabular} 
 \\
 \cline{4-6}
 & \multicolumn{2}{c|}{} & \multicolumn{2}{l|}{\emph{centralized}}  & 
                   \begin{tabular}{p{7.5cm}l} graph coloring~\cite{SagariLondon2015}; cooperation of LTE and \mbox{Wi-Fi}~\cite{Al-Dulaimi2015}
                   \end{tabular}
 \\
\hhline{|~|=====}
 & \multicolumn{4}{l|}{\textbf{Other mechanisms}} & 
                   \begin{tabular}{p{7.5cm}l} opportunistic secondary cell off~\cite{Forum2015}; UL power control~\cite{Abinader2014}; different DL transmit power~\cite{XiaLondon2015, VoicuLondon2015}
                   \end{tabular}
 \\
\hline   
\end{tabular}
\end{center}
\end{table*}

\subsubsection*{Throughput and interference models for a spectrum commons}
Existing models focus on \mbox{Wi-Fi} and follow two main approaches. 
The first approach is the widely used analytical throughput model for IEEE 802.11 CSMA/CA proposed by Bianchi~\cite{Bianchi2000}. However, this model is only applicable to a single specific CSMA/CA technology implementing the same fixed rate at the PHY layer. Most importantly, in~\cite{Bianchi2000} the throughput is modelled only locally, assuming all \mbox{Wi-Fi} devices are within CS range and interference is not taken into account. 
The second approach is based on stochastic geometry models~\cite{H.Q.Nguyen2007} 
that focus on the overall system-level interference bounds, such that \mbox{inter-cell} interference is modelled, but they assume simplified network topologies only and neglect effects of \mbox{environment-specific} propagation conditions. 
An extension to the stochastic geometry models are Monte Carlo simulation-oriented models (e.g.~\cite{VoicuLondon2015}) which additionally take into account individual path losses and interference per device, but still neglect more detailed time-dependent parameters captured by the first approach. 
Importantly, none of these approaches enables comparative evaluation of \mbox{inter-technology} coexistence in a spectrum commons, where CSMA/CA devices coexist with devices implementing a different MAC.  
Our novel throughput and interference model detailed in Section~\ref{section: interf} extends and integrates these approaches, by including the specifics of CSMA/CA coexisting with different variants of duty cycle MAC and \mbox{rate-adaptation} PHY, and by capturing interference with individual path losses and \mbox{long-time} average transmission time of each interferer.     

\subsubsection*{LTE/Wi-Fi coexistence in the unlicensed bands} 
Recent industry and academic research work has proposed several different spectrum sharing mechanisms for LTE in the unlicensed bands, in order to ensure harmonious coexistence with \mbox{Wi-Fi}, as summarized in Table~\ref{table_2}~\cite{Zhang2015, Nhtilae2013, Almeida2013, Jeon2014, Forum2015, Qualcomm2015, Alliance2015, Alliance2016, Zhang2015a, RupasingheNewOrleans2015, Al-Dulaimi2015, 2015, 3GPP2015, BhorkarNewOrleans2015, JiaLondon2015, VoicuLondon2015, ChenGlasgowMay2015, JeonLondon2015, RatasukAustin2014, Song2016, TaoHongKong2015, LiHongKong2015, Bhorkar2014, SagariLondon2015, SallentLondon2015, Abinader2014, XiaLondon2015}. We may classify the spectrum sharing mechanisms as follows: MAC \mbox{time-sharing} mechanisms (i.e. duty cycle, LBT), channel selection mechanisms (i.e. spectrum sharing in frequency), and other mechanisms.  
The performance evaluation in~\cite{Zhang2015, Nhtilae2013, Almeida2013, Jeon2014, Forum2015, Qualcomm2015, Alliance2015, Alliance2016, Zhang2015a, RupasingheNewOrleans2015, Al-Dulaimi2015, 2015, 3GPP2015, BhorkarNewOrleans2015, JiaLondon2015, VoicuLondon2015, ChenGlasgowMay2015, JeonLondon2015, RatasukAustin2014, Song2016, TaoHongKong2015, LiHongKong2015, Bhorkar2014, SagariLondon2015, SallentLondon2015, Abinader2014, XiaLondon2015} of the coexisting LTE variants with \mbox{Wi-Fi} has mostly been done in simplistic scenarios and compared to either original LTE, which does not implement any coexistence mechanism, or with variations of one particular proposed coexistence mechanism. 
In some key works, e.g.~\cite{Forum2015}, the system parameters are incompletely specified, which makes comparison of different spectrum sharing schemes not possible at all. 
The few studies that do consider several coexistence mechanisms mostly evaluate only one MAC approach for LTE in the unlicensed band, i.e. either LBT~\cite{Bhorkar2014, VoicuLondon2015, 3GPP2015} or duty cycle~\cite{Forum2015, Qualcomm2015, Alliance2016, Al-Dulaimi2015}, applied concurrently with channel selection mechanisms, or compare the performance of several channel selection schemes among themselves~\cite{SagariLondon2015}.
To the best of our knowledge, only the authors in~\cite{Jeon2014} consider both MAC coexistence approaches in their study (i.e. LBT and duty cycle), but they only evaluate the fixed duty cycle variant, whereas industry proponents have instead focused their work on the adaptive duty cycle variant for LTE~\cite{Qualcomm2015}.         
Finally, the existing handful of experimental evaluations were either performed with customized, proprietary hardware that is not publicly available~\cite{Qualcomm2015}, or were assessing the coexistence of Wi-Fi with LTE as originally operating in the licensed band, i.e. without any coexistence mechanism~\cite{SagariStockholm2015, JianLondon2015}. 
Unlike previous studies, we select multiple MAC and channel selection techniques proposed in the literature and we vary their key parameters within the same framework, in order to perform a comparative, systematic, and transparent analysis of their behaviour; namely, we do not focus on optimization of parameters under specific and restrictive conditions, thus keeping our analysis and conclusions generalizable.

\section{Spectrum Commons Coexistence Evaluation Methodology}
\label{section: mechanisms}

\subsection{Proposed Framework for Coexistence Evaluation in a Spectrum Commons}
\label{section: mechanisms_1}

We explore the feasible design space for spectrum sharing mechanisms by assuming a population of legacy APs coexisting with a population of new entrant APs. We do so by fixing the technology of the legacy APs and varying key parameters of the spectrum sharing mechanisms implemented by the new entrant APs, which are as follows. 
Spectrum sharing mechanisms facilitate coexistence by partitioning spectrum either in \emph{frequency}, via channel selection, or in \emph{time}, via MAC layer mechanisms, as shown in our classification in Table~\ref{table_2}.
Although both approaches can be applied in a coordinated or a distributed manner, distributed schemes are typically desirable for a spectrum commons, due to the large number of individually managed devices. 

Channel selection schemes reduce interference by assigning different channels to \mbox{co-located} devices, regardless of their level of coordination.
MAC mechanisms for a spectrum commons follow either (i)~distributed sensing approaches, i.e. \emph{\textbf{LBT}}; or (ii)~periodic coordinated/uncoordinated transmission approaches, i.e. \emph{\textbf{duty cycle}}.
For \emph{\textbf{LBT}}, we identify the key parameters, as given in Table~\ref{table_2}, to be: (i)~the CS threshold, which directly affects the tradeoff between sharing the channel in time among multiple APs and suffering from concurrent interference; (ii)~the CS duration (i.e. fixed vs. variable with or without random backoff), which affects the MAC efficiency; and (iii)~the MAC frame duration, which also affects the MAC efficiency. 
For \emph{\textbf{duty cycle}}, the key design parameters in Table~\ref{table_2} are the level of: (i)~adaptiveness when detecting other APs, i.e. \emph{fixed} vs. \emph{adaptive duty cycle}, and (ii)~coordination, i.e. distributed vs. intra- and \mbox{inter-operator} coordination for synchronizing devices and mitigating interference. Additionally, the time granularity (i.e. \mbox{ON-duration}) of the duty cycle can potentially affect the number of collisions with other frames. Further details about the impact of each of these parameters on coexistence are given in Section~\ref{section: interf}.
Although not an intrinsic element of a spectrum sharing mechanism, the PHY layer may interact with the MAC by e.g. changing the coverage area of the APs, affording more robustness to interference, enabling faster/slower frame transmissions, etc.; \mbox{PHY-MAC} interactions should thus also be considered for coexisting AP populations.

Another major system parameter influencing the design of spectrum sharing mechanisms is the \emph{interference coupling} among the devices in the overall heterogeneous network. The interference coupling determines the number of APs within and outside the CS range, the interference from which is handled differently by different MAC mechanisms. In practice, this parameter is determined by the building shielding, the device density, and the transmit power. It is thus imperative to explore a wide range of shielding conditions and device densities, in order to identify the exact cases where a given spectrum sharing mechanism may outperform the others.

\subsection{Case Study Scenarios and Spectrum Sharing Mechanisms}
\label{section: case study variants}

We assume the legacy APs are always IEEE 802.11n \mbox{Wi-Fi} APs implementing CSMA/CA, which is a distributed binary exponential random backoff\footnote{Throughout this paper we consider IEEE 802.11n-like binary exponential random backoff \emph{LBT}.} \emph{LBT} MAC mechanism with \mbox{-82}~dBm CS threshold for deferring to other \mbox{Wi-Fi} devices and \mbox{-62}~dBm CS threshold for deferring to other technologies~\cite{IEEE2012}. 

\begin{table*}[t!]
\caption{New entrant variants for LTE/Wi-Fi coexistence case study in a spectrum commons}
\label{table_entrant}
\centering
\begin{tabular}{|c|l|l|}
\hline
\textbf{PHY layer} & \multicolumn{1}{c|}{\textbf{MAC layer}} & \multicolumn{1}{c|}{\textbf{Comments}} \\
\hline
IEEE 802.11n & LBT with -82~dBm CS threshold & \begin{tabular}{l} IEEE 802.11n \mbox{Wi-Fi} \end{tabular}\\
\hline
IEEE 802.11n & LBT with -62~dBm CS threshold & \begin{tabular}{l} \mbox{Wi-Fi} with higher CS threshold \end{tabular}\\
\hline
LTE & always on & \begin{tabular}{l} standard LTE (no time sharing) \end{tabular}\\
\hline
LTE & LBT with -62~dBm CS threshold & \begin{tabular}{l} \mbox{Wi-Fi-like} LBT \end{tabular}\\
\hline
LTE & fixed 50\% coordinated duty cycle & \begin{tabular}{l} local coordination, such that all transmissions \\within CS range (-62~dBm) overlap in time\end{tabular}\\
\hline
LTE & fixed 50\% uncoordinated duty cycle & \begin{tabular}{l} random transmissions in time \end{tabular}\\
\hline
LTE & adaptive duty cycle & \begin{tabular}{l} random transmissions in time; adaptation based on \\number of APs detected with -62~dBm CS threshold, \\to be comparable with \emph{LBT} \end{tabular}\\
\hline
LTE & ideal TDMA & \begin{tabular}{l} adaptive duty cycle with perfect local coordination, \\such that no transmissions within CS range overlap in time\end{tabular} \\
\hline
\end{tabular}
\end{table*}

The new entrant APs represent candidate LTE technologies for the unlicensed band, with the combinations of PHY and MAC layers given in Table~\ref{table_entrant}, reflecting the classification of coexistence mechanism design parameters in Section~\ref{section: mechanisms_1}. The first two entrant variants implement the IEEE 802.11n PHY and are considered as the baseline reference, as they represent IEEE 802.11n \mbox{Wi-Fi} devices with different CS thresholds. The remaining entrant variants implement the LTE PHY and we vary their MAC mechanisms, as follows: \emph{LBT}, \emph{fixed 50\% duty cycle} (\emph{coordinated} and \emph{uncoordinated}), and \emph{adaptive duty cycle}. As boundary cases for interference management, we consider the \emph{always on} MAC, where all APs transmit continuously (i.e. highest interference), and the \emph{ideal TDMA} MAC, which essentially is an \emph{adaptive duty cycle} with perfect local coordination (i.e. lowest interference). We note that the level of coordination assumed for \emph{ideal TDMA} would require in practice perfect intra- and inter-operator coordination.     

We consider a total number of 19 indoor and 11 outdoor 20~MHz channels in the 5~GHz unlicensed band~\cite{IEEE2012}. We assume legacy APs randomly select one channel per AP, whereas new entrant APs either apply \emph{\textbf{random}} channel selection, or \emph{\textbf{sense}} channel selection, where each entrant AP randomly selects a channel that is not occupied by legacy APs. We also consider the \emph{\textbf{single channel}} scheme, with only \mbox{co-channel} legacy and entrant APs, which enables a detailed investigation of spectrum sharing mechanisms in time.

\begin{figure}[t!]
\centering
\subfigure[\emph{Indoor/indoor}: building with  $10\times10\times3$~m apartments and a 5~m margin around the building.]
          {\includegraphics[width=0.9\linewidth]{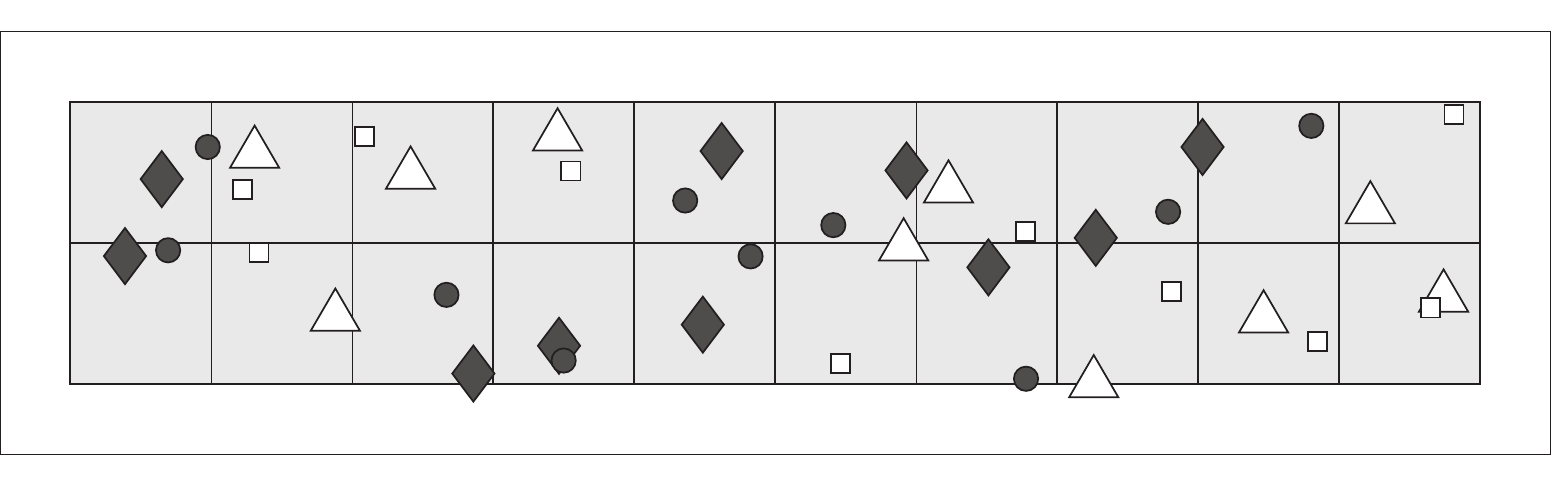}
             \label{fig_layout_ii}}
\\
\subfigure[\emph{Indoor/outdoor}: random length and height of buildings between 3-10 apartments and 3-5 floors, respectively; total area of $346\times389$~m.]
          {\includegraphics[width=0.82\linewidth]{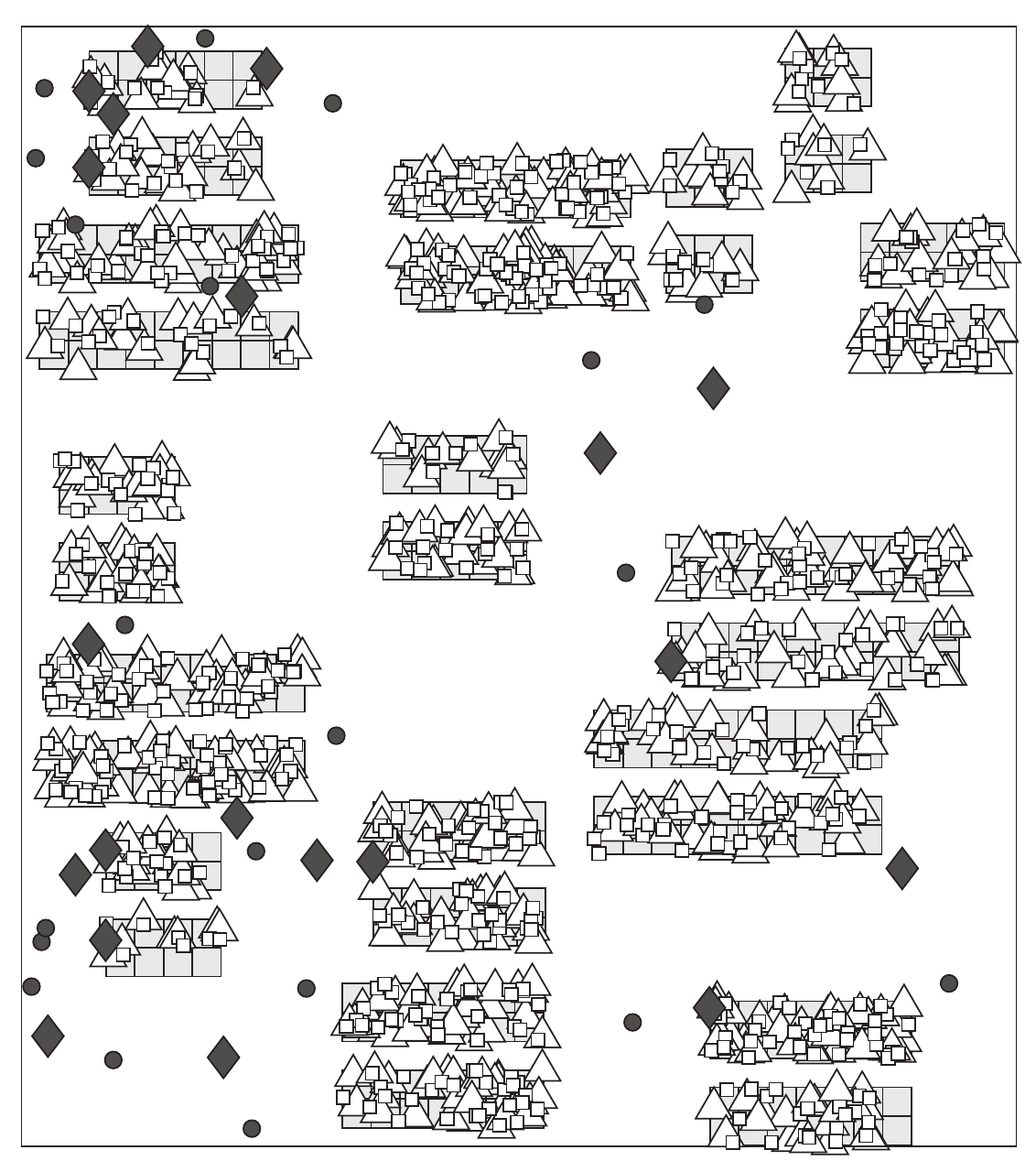}
           \label{fig_layout_oi}}
\\
\subfigure[\emph{Outdoor/outdoor}.]      	  	 
            {\includegraphics[width=0.82\linewidth]{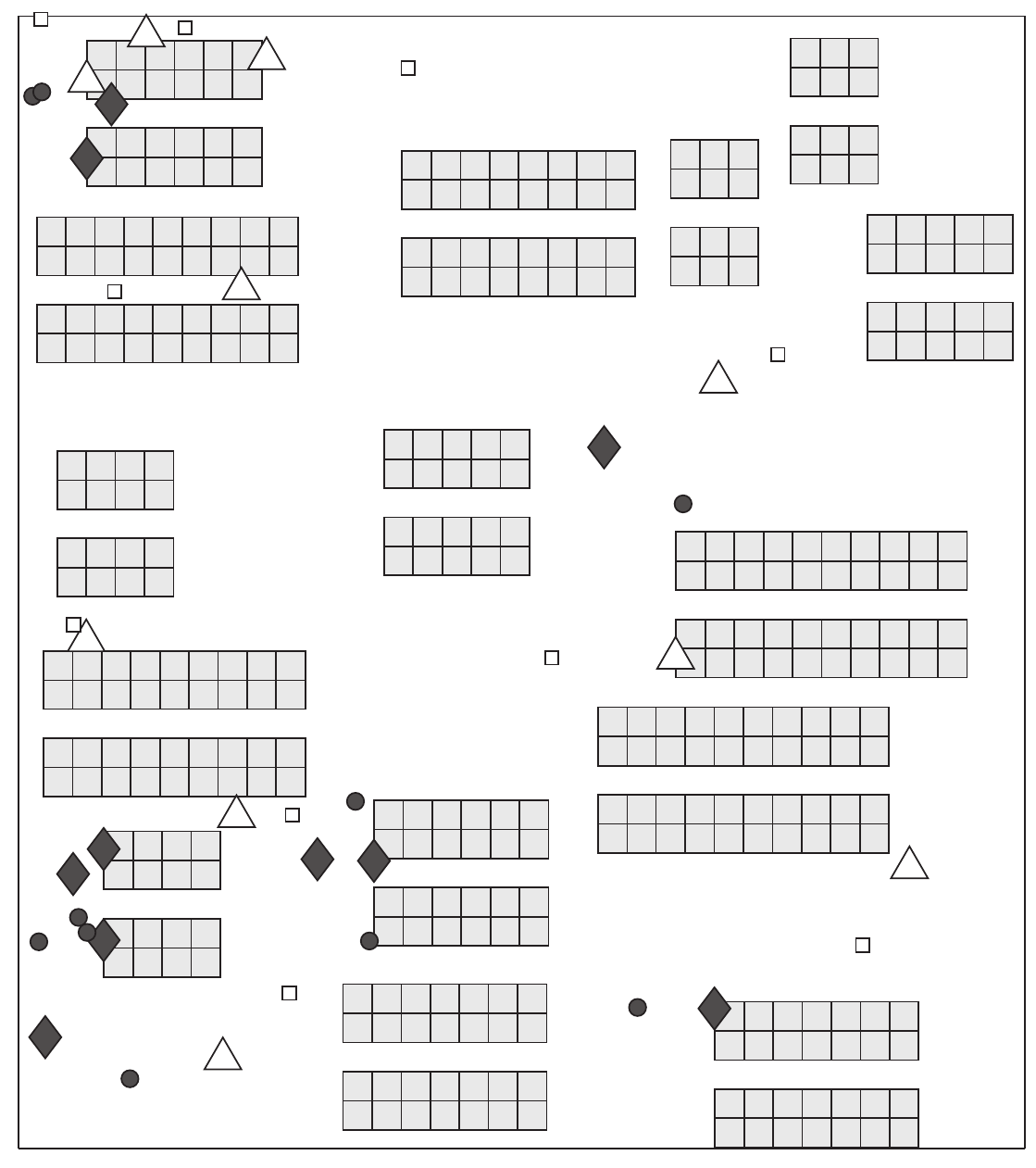}
            \label{fig_layout_oo}}
\caption{Example top-view network layout for (a) \emph{indoor/indoor}, (b) \emph{indoor/outdoor}, and (c) \emph{outdoor/outdoor} scenarios, showing legacy APs ($\bigtriangleup$), legacy users ($\Box$), new entrant APs ($\color{black} \blacklozenge$), and new entrant users ($\bullet$).}
\label{fig_layout}
\end{figure}

In contrast to existing studies which consider coexistence in restricted scenarios~\cite{Zhang2015, Nhtilae2013, Almeida2013, Jeon2014, Forum2015, Qualcomm2015, Alliance2015, Alliance2016, Zhang2015a, RupasingheNewOrleans2015, Al-Dulaimi2015, 2015, 3GPP2015, BhorkarNewOrleans2015, JiaLondon2015, VoicuLondon2015, ChenGlasgowMay2015, JeonLondon2015, RatasukAustin2014, Song2016, TaoHongKong2015, LiHongKong2015, Bhorkar2014, SagariLondon2015, SallentLondon2015, Abinader2014, XiaLondon2015}, 
we explore a wide and realistic range of interference coupling conditions, by varying the legacy and new entrant AP densities in four deployment scenarios. These scenarios are based on real outdoor base station locations and the 3GPP dual stripe model in~\cite{Alcatel-Lucent2009}, as follows.
For the \textbf{\emph{indoor/indoor} scenario}, both legacy and new entrant APs are located indoors. We assume one \mbox{single-floor} \mbox{dual-stripe} building, where each AP and its associated user are randomly located in a single apartment, as shown in Fig.~\ref{fig_layout_ii}. We note that throughout this paper we assume each AP has one associated user. We consider either 1 or 10 legacy APs and 1 to 10 new entrant APs. The equivalent overall network density is thus 600-6000~APs/km\textsuperscript{2}, consistent with recent \mbox{Wi-Fi} measurements in~\cite{Achtzehn2013}. 
We also study the \textbf{\emph{indoor/indoor}} \textbf{scenario} \textbf{\emph{without internal walls}} as a variant of the \emph{indoor/indoor} scenario above, since this gives a lower bound for wall shielding, which would lead to increased interference and thus to a more congested coexistence scenario.
For the \textbf{\emph{indoor/outdoor} scenario}, the legacy APs are located indoors and the new entrant APs are located outdoors. For the outdoor APs we consider real base station locations obtained through measurements in central London~\cite{MozillaLocationService2015}. 
Out of all locations in~\cite{MozillaLocationService2015} we select 20 locations, which are representative for low transmit power entrant APs, with a coverage range of up to 300~m and with minimum 20 measurement observations. We assume the height of the outdoor APs is at the building roof level. The associated entrant users are randomly located outdoors, in the coverage area of the respective entrant APs that they are associated with, at a maximum distance of 50~m from the respective APs, and at a height of 1.5~m. We overlay randomly located buildings on the study area where the outdoor APs are distributed, as shown in Fig.~\ref{fig_layout_oi}. The indoor APs and users are randomly located in apartments with a density of 500 or 5000~APs/km\textsuperscript{2}~\cite{Achtzehn2013}. We vary the number of outdoor APs from 1 to 20 (equivalent entrant densities of \mbox{7-150}~APs/km\textsuperscript{2}).
For the \textbf{\emph{outdoor/outdoor} scenario}, both legacy and entrant APs are located outdoors. We assume a similar network layout as for the \emph{indoor/outdoor} scenario, where the outdoor locations are randomly assigned to legacy and entrant APs, as shown in Fig.~\ref{fig_layout_oo}. 
In accordance with regulatory limits, we assume indoor legacy and new entrant APs transmit with a power level of 23~dBm, whereas outdoor legacy and new entrant APs transmit with a power level of 30~dBm. 

\section{Throughput and Interference Model for Heterogeneous Devices Coexisting in a Spectrum Commons}
\label{section: interf}

We propose a novel integrated \mbox{inter-technology} throughput and interference model that incorporates different spectrum sharing mechanisms and their key parameters at the same level of abstraction, for populations of APs in large-scale networks, with overlapping or \mbox{non-overlapping} coverage areas of individual APs. In this section we first present the general formulation of our model and then apply it to our LTE/\mbox{Wi-Fi} case study.
Without loss of generality\footnote{Importantly, although we focus here on the \mbox{two-technology} coexistence case, our model can be straightforwardly reduced to the \mbox{single-technology} case, or extended to the \mbox{multi-technology} coexistence case in a spectrum commons.}, we assume a population $\mathbf{M}$ of same technology APs that share a channel with another population of APs $\mathbf{N}$. For our case study, we use the model for populations $\mathbf{M}$ and $\mathbf{N}$ for the legacy and new entrant APs, respectively, with the respective combinations of MAC and PHY specified in Section~\ref{section: case study variants}.  
We always assume downlink saturated traffic and only one user per AP, at which we estimate the throughput\footnote{For multiple users per AP, the throughput per user would simply be a fraction of our estimated throughput per AP.}. 
A \mbox{high-level} description of our model, with respect to the spectrum sharing mechanisms given in Section~\ref{section: mechanisms}, is as follows.

\begin{figure}[t!]
\centering
\subfigure[CS range.]
          {\includegraphics[scale=0.25]{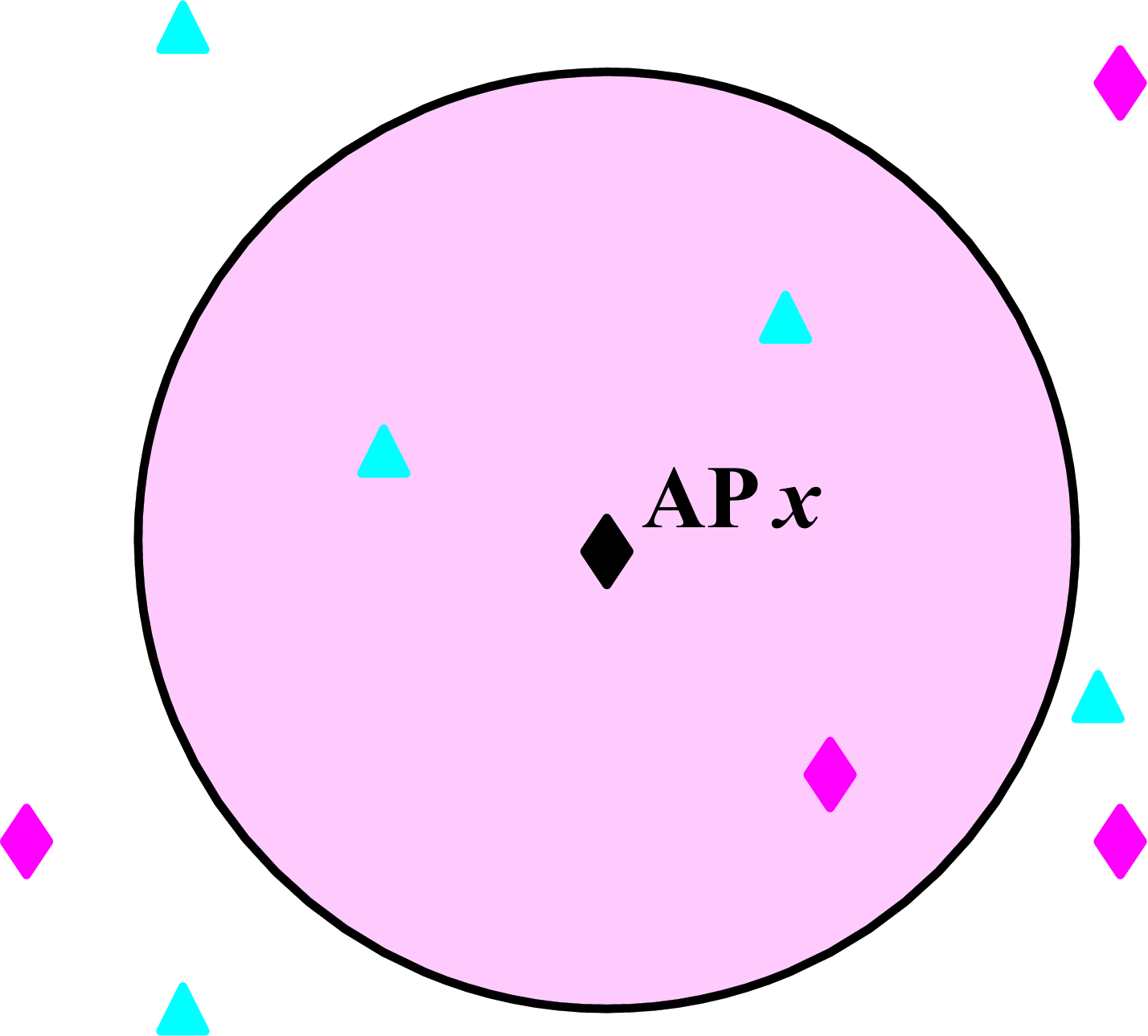}
           \label{fig_x_1}}
\\
\subfigure[Channel time fraction for AP $x$ in (a), in gray.]      	
	   	{\includegraphics[width=\linewidth]{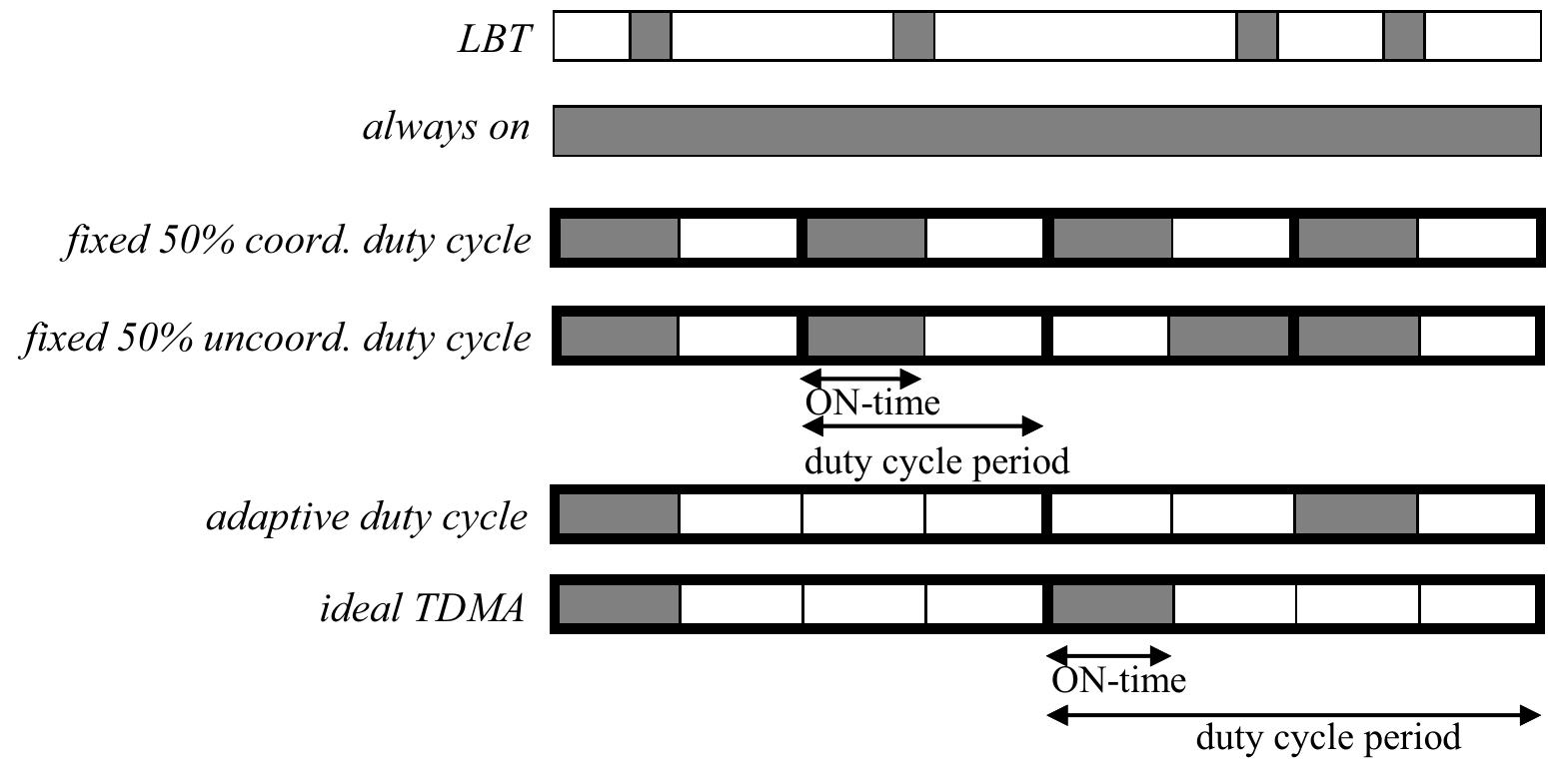}
		   \label{fig_x_2}}
\caption{Illustration of (a) CS range containing 3 other APs, for AP $x$ ($\color{black} \blacklozenge$) in $\mathbf{N}$, when AP populations $\mathbf{M}$ ($\color{cyan} \blacktriangle$) and $\mathbf{N}$ ($\color{magenta} \blacklozenge$) coexist on the same channel and (b) how coexistence is managed for AP $x$ by the considered MAC mechanisms in Table~\ref{table_entrant}.}
\label{fig_x}
\end{figure}

For \emph{LBT}, we assume all co-channel APs within each other's CS range are prevented from transmitting simultaneously. We assume an AP implementing \emph{LBT} occupies the channel only for a fraction of time roughly equal to the inverse of the sum of the number of \mbox{co-channel} APs in its CS range~\cite{LjiljanaSimic2012, H.Q.Nguyen2007}, while the rest of the time is used by other APs in its CS range, and we estimate the downlink throughput per AP accordingly. Co-channel APs located outside the CS range of the considered AP interfere with this AP by decreasing the \mbox{signal-to-interference-and-noise}~ratio (SINR) of its associated user. Fig.~\ref{fig_x_1} shows an example of the CS range for one AP when populations $\mathbf{M}$ and $\mathbf{N}$ coexist. 

\begin{table*}[t!]
\begin{center}
\caption{Throughput parameters for AP $x$ from $\mathbf{M}$ and AP $y$ from $\mathbf{N}$ coexisting on the same channel \label{table_4} }
\begin{tabular}{|l|c|c|c|c|}
\hline
	\diagbox[width=2.5cm]{\textbf{Parameters}}{\textbf{MAC} \\ ($\mathbf{M}$, $\mathbf{N}$)} 
	& \begin{tabular}{l} $\mathbf{M}$: LBT \\ $\mathbf{N}$: LBT \end{tabular}
	& \begin{tabular}{l} $\mathbf{M}$: LBT \\ $\mathbf{N}$: always on \end{tabular}
	& \begin{tabular}{l} $\mathbf{M}$: LBT \\ $\mathbf{N}$: adaptive duty cycle/\\ideal TDMA \end{tabular}
	& \begin{tabular}{l} $\mathbf{M}$: LBT \\ $\mathbf{N}$: fixed 50\% duty cycle \end{tabular} \\
\hline
\hline
	$S^M_x$	& $S_x$, Section~\ref{section: MAC OH} & $S_x$, Section~\ref{section: MAC OH} & $S_x$, Section~\ref{section: MAC OH} & $S_x$, Section~\ref{section: MAC OH} \\
\hline
	$S^N_y$ & $S_x$, Section~\ref{section: MAC OH} & 1 & 1 & 1 \\
\hline
\hline
    $COLL^M_x$ & 1 & 1 & $(1-r^{M,N}_{deg}(x))$, Section~\ref{section: col model} & $(1-r^{M,N}_{deg}(x))$, Section~\ref{section: col model} \\
\hline
    $COLL^N_y$ & 1 & 1 & 1 & 1 \\	
\hline
\hline
    $AirTime^M_x$ & $\frac{1}{1+|\mathbf{A}_x|+|\mathbf{B}_x|}$ & $f^{M,N}_{dut}(x) \times \frac{1}{1+|\mathbf{A}_x|}$ & $f^{M,N}_{dut}(x) \times \frac{1}{1+|\mathbf{A}_x|}$ & $f^{M,N}_{dut}(x) \times \frac{1}{1+|\mathbf{A}_x|}$ \\
\hline
    $AirTime^N_y$ & $\frac{1}{1+|\mathbf{C}_y|+|\mathbf{D}_y|}$ & 1 & $\frac{1}{1+|\mathbf{C}_y|+|\mathbf{D}_y|}$ & $\frac{1}{2}$ \\
\hline
\end{tabular}
\end{center}
\end{table*}

\begin{table}[t!]
\begin{center}
\caption{Notation \label{table_1} }
\begin{tabular}{|l|l|}
\hline
	$\mathbf{A}$ & \begin{tabular}{l} the set of all \mbox{co-channel} APs in $\mathbf{M}$ \end{tabular}\\
\hline
	$\mathbf{A}_x$ & \begin{tabular}{l} the set of \mbox{co-channel} APs in $\mathbf{M}$ \\and in the CS range of AP $x$ in $\mathbf{M}$ \end{tabular}\\
\hline
    $|\mathbf{A}_x|$ & \begin{tabular}{l} the number of \mbox{co-channel} APs in $\mathbf{M}$ \\and in the CS range of AP $x$ in $\mathbf{M}$ \end{tabular}\\
\hline
    $\mathbf{B}$ & \begin{tabular}{l} the set of all \mbox{co-channel} APs in $\mathbf{N}$ \end{tabular}\\
\hline
    $\mathbf{B}_x$	& \begin{tabular}{l} the set of \mbox{co-channel} APs in $\mathbf{N}$ \\and in the CS range of AP $x$ in $\mathbf{M}$ \end{tabular}\\
\hline
    $|\mathbf{B}_x|$ & \begin{tabular}{l} the number of \mbox{co-channel} APs in $\mathbf{N}$ \\and in the CS range of AP $x$ in $\mathbf{M}$ \end{tabular}\\
\hline
    $\mathbf{C}_y$	& \begin{tabular}{l} the set of \mbox{co-channel} APs in $\mathbf{M}$ \\and in the CS range of AP $y$ in $\mathbf{N}$ \end{tabular}\\
\hline
    $|\mathbf{C}_y|$ & \begin{tabular}{l} the number of \mbox{co-channel} APs in $\mathbf{M}$ \\and in the CS range of AP $y$ in $\mathbf{N}$ \end{tabular}\\
\hline
    $\mathbf{D}_y$	& \begin{tabular}{l} the set of \mbox{co-channel} APs in $\mathbf{N}$ \\and in the CS range of AP $y$ in $\mathbf{N}$ \end{tabular}\\
\hline
    $|\mathbf{D}_y|$ & \begin{tabular}{l} the number of \mbox{co-channel} APs in $\mathbf{N}$ \\and in the CS range of AP $y$ in $\mathbf{N}$ \end{tabular}\\
\hline
    $P^M$	& \begin{tabular}{l} transmit power of an AP in $\mathbf{M}$ \end{tabular}\\
\hline
	$P^N$	& \begin{tabular}{l} transmit power of an AP in $\mathbf{N}$  \end{tabular}\\
\hline
    $L_{u,x}$ ($L_{v,y}$) & \begin{tabular}{l} path loss between user $u$ ($v$) \\and its associated AP $x$ ($y$)  \end{tabular}\\
\hline 
    $L_{u,z}$ ($L_{v,z}$) & \begin{tabular}{l} path loss between user $u$ ($v$) and AP $z$  \end{tabular}\\     
\hline
    $I^M_u$ ($I^M_v$) & \begin{tabular}{l} the aggregated co-channel interference \\at user $u$ ($v$) from AP population $\mathbf{M}$ \end{tabular}\\
\hline 
    $I^N_u$ ($I^N_v$) & \begin{tabular}{l} the aggregated co-channel interference \\at user $u$ ($v$) from AP population $\mathbf{N}$ \end{tabular}\\        
\hline
    $N_0$  & \begin{tabular}{l} noise power (-174~dBm/Hz)  \end{tabular} \\
\hline
\end{tabular}
\end{center}
\end{table}

For \emph{duty cycle}, without loss of generality, we adopt a distributed time-slotted model to simplify our analysis\footnote{We note that the slotted model gives the same \emph{long-term average} throughput statistics per entrant/legacy AP as would be observed for the general case of fully distributed asynchronous devices (i.e. non-slotted transmissions, whereby devices are not restricted to starting transmissions only at the start of a time slot). In the non-slotted model, different entrant APs would have random partially overlapping transmissions, which \emph{in the long-term} results in the same level of mutual interference as does the random occurrence of either fully-overlapping or non-overlapping transmissions in the slotted model. Similarly, in the non-slotted model legacy Wi-Fi APs would find the channel unoccupied by entrants for an \emph{equivalent long-term fraction of time}, resulting in the same channel share for legacy APs as in the slotted model (where the channel is unoccupied by entrants for random time durations which are simply a \emph{multiple} of time slots).
}.
We assume all APs use the same time slot
duration, as illustrated in Fig.~\ref{fig_x_2}. For \emph{fixed 50\% coordinated duty cycle} all APs transmit in the same time slot, whereas for \emph{adaptive duty cycle} and \emph{fixed 50\% uncoordinated duty cycle} each AP randomly selects a time slot to transmit in, such that transmissions from different co-channel APs may or may not overlap in the same time slot. 
For both variations of \emph{fixed 50\% coordinated/uncoordinated duty cycle}, the total duration of a time period is 2 time slots for all APs, whereas for \emph{adaptive duty cycle} each AP calculates its own duty cycle period as the number of time slots equal to the number of APs in its CS range. Consequently, the throughput of each AP is proportional to its duty cycle and the SINR of its associated user is decreased by the interference from all other co-channel APs, according to their own transmission time. We model the throughput of \emph{ideal TDMA} as being that of perfectly coordinated \emph{adaptive duty cycle}, i.e. without additional control overhead.

In the remainder of this section we firstly present the throughput model, then the interference and SINR model. Finally, we present details of the \emph{LBT} MAC overhead and we model the throughput degradation due to frame collisions when \emph{duty cycle} devices coexist with \emph{LBT} devices.

\subsection{Throughput Model}

In general, we model the throughput of AP $x$ from population $\mathbf{M}$ as 
\begin{equation}\label{eq_28}
R^M_x=S^M_x \times COLL^{M}_x \times AirTime^M_x \times \rho^M_x(SINR^M_u),
\end{equation}
where $S^M_x$ is the MAC efficiency of $x$, $COLL^{M}_x$ is the throughput degradation of $x$ due to collisions between its frames and frames from population $\mathbf{N}$ -- if $\mathbf{N}$ implements duty cycle, $AirTime^M_x$ is the fraction of time that $x$ obtains according to its MAC and the MAC of other APs in its CS range, and $\rho^M_x(SINR^M_u)$ is an \mbox{auto-rate} function mapping the $SINR$ of the associated user $u$ to the PHY spectral efficiency. For our case study we use the example \mbox{auto-rate} functions of IEEE 802.11n~\cite{IEEE2012} and LTE~\cite{3GPP2009}. For LTE PHY we assume the noise figure NF=9~dB~\cite{3GPP2009}, whereas for IEEE 802.11n PHY, NF=15~dB~\cite{IEEE2012}. The throughput $R^N_y$ of an AP $y$ from population $\mathbf{N}$ is expressed analogously. 
The parameters in~(\ref{eq_28}) are specified in Table~\ref{table_4}, for different combinations of coexisting MAC mechanisms, where we use the definitions in Table~\ref{table_1} and $f^{M,N}_{dut}(x)$ is the probability that a duty cycle time slot within a period is not used by any AP from population $\mathbf{N}$ within the CS range of AP $x$ from population $\mathbf{M}$, given in~(\ref{eq_32}). 


We note that for \emph{LBT}, populations $\mathbf{M}$ and $\mathbf{N}$ may each have a different CS threshold to define their CS range. The MAC efficiency, $S^M_x$ and $S^N_y$ (detailed in Section~\ref{section: MAC OH}), is lower than 1 only for \emph{LBT} which wastes transmission time due to the sensing duration. If there is an AP in $\mathbf{N}$ implementing \emph{always on} within the CS range of an AP $x$ in $\mathbf{M}$ implementing \emph{LBT}, we assume $R^M_x$=0, since $AirTime^M_x$=0. We assume that during a collision between \emph{LBT} and \emph{duty cycle} frames only the \emph{LBT} frame is completely lost; we discuss this in more detail in Section~\ref{section: col model}. If all APs in $\mathbf{N}$, that are in the CS range of AP $x$ in $\mathbf{M}$, transmit in the same time slot (i.e. \emph{fixed 50\% coordinated duty cycle}), the other time slot in the time period is shared in time between those co-channel APs in $\mathbf{M}$ that are in the CS range of AP $x$. If each AP in $\mathbf{N}$ randomly selects one of the two time slots to transmit in (i.e. \emph{fixed 50\% uncoordinated duty cycle}), we calculate a long-time average of the fraction of time slots that are unoccupied by co-channel APs in $\mathbf{N}$ within one time period, assuming that each  AP in $\mathbf{N}$, that is in the CS range of AP $x$ in $\mathbf{M}$, selects any of the two time slots with probability $\frac{1}{2}$. If the APs in $\mathbf{N}$ implement \emph{adaptive duty cycle}, each AP in $\mathbf{N}$ calculates its own number of time slots in a time period and randomly selects one time slot to transmit in, in each period. Again we calculate a long-time average of the fraction of time slots that are unoccupied by APs in $\mathbf{N}$.

\begin{figure*}[!t]
\normalsize
\setcounter{MYtempeqncnt}{\value{equation}}
\setcounter{equation}{1}
\begin{equation}\label{eq_32}
f^{M,N}_{dut}(x)=\begin{cases}
	\frac{1}{2}, & \text{if $\mathbf{N}$ has \emph{50\% coordinated duty cycle} \& } \mathbf{B}_x \neq \varnothing \\
	\big(1-\frac{1}{2}\big)^{|\mathbf{B}_x|}, & \text{if $\mathbf{N}$ has \emph{50\% uncoordinated duty cycle} \& } \mathbf{B}_x \neq \varnothing \\
	\displaystyle \prod_{y\in{\mathbf{B}_x}} \Big(1-\frac{1}{1+|\mathbf{C}_y|+|\mathbf{D}_y|}\Big), & \text{if $\mathbf{N}$ has \emph{adaptive duty cycle} \& } \mathbf{B}_x \neq \varnothing \\
	\frac{1+|\mathbf{A}_x|}{1+|\mathbf{A}_x|+|\mathbf{B}_x|}, & \text{if $\mathbf{N}$ has \emph{ideal TDMA} \& } \mathbf{B}_x \neq \varnothing\\
	0, & \text{if $\mathbf{N}$ has \emph{always on} \& } \mathbf{B}_x \neq \varnothing \\
	1, & \text{if } \mathbf{B}_x=\varnothing
\end{cases}
\end{equation} 
\setcounter{equation}{\value{MYtempeqncnt}}
\hrulefill
\vspace*{4pt}
\end{figure*}

\addtocounter{equation}{1}

\subsection{Interference and SINR Model}

The SINR of user $u$ associated with AP $x$ in $\mathbf{M}$ coexisting with $\mathbf{N}$ is given by
\begin{equation}\label{eq_29}
SINR^{M}_u=\frac{P^M(L_{u,x})^{-1}}{N_0+ I^M_u+ I^N_u},
\end{equation}
where we assume the definitions in Table~\ref{table_1}. The SINR of a user $v$ associated with AP $y$ in $\mathbf{N}$ is expressed analogously. We note that the interference term $I^M_u+ I^N_u$ depends on the MAC mechanisms (i.e. air time) of the \mbox{co-channel} interfering APs.

Tables~\ref{table_6} and~\ref{table_7} give the interference parameters in~(\ref{eq_29}) for user $u$ associated with AP $x$ in $\mathbf{M}$ and for user $v$ associated with AP $y$ in $\mathbf{N}$, respectively. 
We note that $I^M_u$ and $I^M_v$ are similar, as both represent \mbox{outside-CS-range} interference from an \emph{LBT} population, which avoids interference within the CS range, irrespective of the coexisting population. Also, local coordination within the CS range is done differently for \emph{fixed 50\% coordinated duty cycle} compared to \emph{adaptive duty cycle}: for \emph{fixed 50\% coordinated duty cycle} the APs within the CS range always transmit at the same time, therefore interference is increased, whereas for \emph{ideal TDMA} (i.e. coordinated \emph{adaptive duty cycle}) interference is completely eliminated within the CS range. For both uncoordinated versions of these respective duty cycle approaches, interference within CS range is randomized. Interference from outside the CS range is randomized for all MAC mechanisms.

\begin{table*}[t!]
\begin{center}
\caption{Interference model for user $u$ associated with AP $x$ from $\mathbf{M}$ coexisting with $\mathbf{N}$ on the same channel \label{table_6} }
\begin{tabular}{|l|c|c|}
\hline
	\diagbox{\textbf{MAC for} $\mathbf{M}$ \textbf{and} $\mathbf{N}$}{\textbf{Parameters}}	& $I^M_u$ & $I^N_u$ \\
\hline
	\begin{tabular}{l} $\mathbf{M}$: LBT, \\$\mathbf{N}$: LBT \end{tabular} 
	& $\displaystyle \sum_{z\in{\mathbf{A}\smallsetminus\mathbf{A}_x}} \frac{P^M(L_{u,z})^{-1}}{1+|\mathbf{A}_z|+|\mathbf{B}_z|} $ 
	& $\displaystyle \sum_{z\in{\mathbf{B}\smallsetminus\mathbf{B}_x}} \frac{P^N(L_{u,z})^{-1}}{1+|\mathbf{C}_z|+|\mathbf{D}_z|}$\\
\hline
    \begin{tabular}{l} $\mathbf{M}$: LBT, \\$\mathbf{N}$: always on \end{tabular} 
    & $\displaystyle \sum_{z\in{\mathbf{A}\smallsetminus\mathbf{A}_x}} f^{M,N}_{dut}(z) \times \frac{P^M(L_{u,z})^{-1}}{1+|\mathbf{A}_z|} $
    & $\displaystyle \sum_{z\in{\mathbf{B}\smallsetminus\mathbf{B}_x}} P^N(L_{u,z})^{-1}$  \\
\hline
    \begin{tabular}{l} $\mathbf{M}$: LBT, \\$\mathbf{N}$: adaptive duty cycle \end{tabular} 
    & $\displaystyle \sum_{z\in{\mathbf{A}\smallsetminus\mathbf{A}_x}} f^{M,N}_{dut}(z) \times \frac{P^M(L_{u,z})^{-1}}{1+|\mathbf{A}_z|} $
    & $\displaystyle \sum_{z\in{\mathbf{B}\smallsetminus\mathbf{B}_x}} \frac{P^N(L_{u,z})^{-1}}{1+|\mathbf{C}_z|+|\mathbf{D}_z|}$ \\
\hline
    \begin{tabular}{l} $\mathbf{M}$: LBT, \\$\mathbf{N}$: ideal TDMA \end{tabular} 
    & $\displaystyle \sum_{z\in{\mathbf{A}\smallsetminus\mathbf{A}_x}} \frac{P^M(L_{u,z})^{-1}}{1+|\mathbf{A}_z|+|\mathbf{B}_z|} $ 
	& $\displaystyle \sum_{z\in{\mathbf{B}\smallsetminus\mathbf{B}_x}} \frac{P^N(L_{u,z})^{-1}}{1+|\mathbf{C}_z|+|\mathbf{D}_z|}$  \\
\hline 
    \begin{tabular}{l} $\mathbf{M}$: LBT, \\$\mathbf{N}$: fixed 50\% coordinated duty cycle \end{tabular} 
    & $\displaystyle \sum_{z\in{\mathbf{A}\smallsetminus\mathbf{A}_x}} {f^{M,N}_{dut}(z)\times \frac{P^M(L_{u,z})^{-1}} {1+|\mathbf{A}_z|}} $
    & $\displaystyle \sum_{z\in{\mathbf{B}\smallsetminus\mathbf{B}_x}} \frac{P^N(L_{u,z})^{-1}}{2}$ \\
\hline
    \begin{tabular}{l} $\mathbf{M}$: LBT, \\$\mathbf{N}$: fixed 50\% uncoordinated duty cycle \end{tabular} 
    & $\displaystyle \sum_{z\in{\mathbf{A}\smallsetminus\mathbf{A}_x}} f^{M,N}_{dut}(z) \times \frac{P^M(L_{u,z})^{-1}}{1+|\mathbf{A}_z|} $
    & $\displaystyle \sum_{z\in{\mathbf{B}\smallsetminus\mathbf{B}_x}} \frac{P^N(L_{u,z})^{-1}}{2}$  \\            
\hline
\end{tabular}
\end{center}
\end{table*}

\begin{table*}[t!]
\begin{center}
\caption{Interference model for user $v$ associated with AP $y$ from $\mathbf{N}$ coexisting with $\mathbf{M}$ on the same channel \label{table_7} }
\begin{tabular}{|l|c|c|}
\hline
	\diagbox{\textbf{MAC} ($\mathbf{M}$ \textbf{and} $\mathbf{N}$)}{\textbf{Parameters}}	& $I^M_v$ & $I^N_v$ \\
\hline
	\begin{tabular}{l} $\mathbf{M}$: LBT, \\$\mathbf{N}$: LBT \end{tabular} 
	& $\displaystyle \sum_{z\in{\mathbf{A}\smallsetminus\mathbf{C}_y}} \frac{P^M(L_{v,z})^{-1}}{1+|\mathbf{A}_z|+|\mathbf{B}_z|} $ 
	& $\displaystyle \sum_{z\in{\mathbf{B}\smallsetminus\mathbf{D}_y}} \frac{P^N(L_{v,z})^{-1}}{1+|\mathbf{C}_z|+|\mathbf{D}_z|}$\\
\hline
    \begin{tabular}{l} $\mathbf{M}$: LBT, \\$\mathbf{N}$: always on \end{tabular} 
    & $\displaystyle \sum_{z\in{\mathbf{A}\smallsetminus\mathbf{C}_y}} f^{M,N}_{dut}(z)\times \frac{P^M(L_{v,z})^{-1}}{1+|\mathbf{A}_z|}$
    & $\displaystyle \sum_{z\in{\mathbf{D}_y}} P^N(L_{v,z})^{-1} +\sum_{z\in{\mathbf{B}\smallsetminus\mathbf{D}_y}} P^N(L_{v,z})^{-1}$  \\
\hline
    \begin{tabular}{l} $\mathbf{M}$: LBT, \\$\mathbf{N}$: adaptive duty cycle \end{tabular} 
    & $\displaystyle \sum_{z\in{\mathbf{A}\smallsetminus\mathbf{C}_y}} f^{M,N}_{dut}(z)\times \frac{P^M(L_{v,z})^{-1}}{1+|\mathbf{A}_z|} $
    & $\displaystyle \sum_{z\in{\mathbf{D}_y}}\frac{P^N(L_{v,z})^{-1}}{1+|\mathbf{C}_z|+|\mathbf{D}_z|} +\sum_{z\in{\mathbf{B}\smallsetminus\mathbf{D}_y}} \frac{P^N(L_{v,z})^{-1}}{1+|\mathbf{C}_z|+|\mathbf{D}_z|}$ \\
\hline
    \begin{tabular}{l} $\mathbf{M}$: LBT, \\$\mathbf{N}$: ideal TDMA \end{tabular} 
    & $\displaystyle \sum_{z\in{\mathbf{A}\smallsetminus\mathbf{C}_y}} \frac{P^M(L_{v,z})^{-1}}{1+|\mathbf{A}_z|+|\mathbf{B}_z|} $ 
	& $\displaystyle \sum_{z\in{\mathbf{B}\smallsetminus\mathbf{D}_y}} \frac{P^N(L_{v,z})^{-1}}{1+|\mathbf{C}_z|+|\mathbf{D}_z|}$ \\
\hline 
    \begin{tabular}{l} $\mathbf{M}$: LBT, \\$\mathbf{N}$: fixed 50\% coord. duty cycle \end{tabular}
    & $\displaystyle \sum_{z\in{\mathbf{A}\smallsetminus\mathbf{C}_y}} f^{M,N}_{dut}(z)\times \frac{P^M(L_{v,z})^{-1}}{1+|\mathbf{A}_z|} $ 
    & $\displaystyle \sum_{z\in{\mathbf{D}_y}}P^N(L_{v,z})^{-1} +\sum_{z\in{\mathbf{B}\smallsetminus\mathbf{D}_y}} \frac{P^N(L_{v,z})^{-1}}{2}$  \\
\hline
    \begin{tabular}{l} $\mathbf{M}$: LBT, \\$\mathbf{N}$: fixed 50\% uncoord. duty cycle \end{tabular}
    & $\displaystyle \sum_{z\in{\mathbf{A}\smallsetminus\mathbf{C}_y}} f^{M,N}_{dut}(z)\times \frac{P^M(L_{v,z})^{-1}}{1+|\mathbf{A}_z|} $
    & $\displaystyle \sum_{z\in{\mathbf{D}_y}}\frac{P^N(L_{v,z})^{-1}}{2} +\sum_{z\in{\mathbf{B}\smallsetminus\mathbf{D}_y}} \frac{P^N(L_{v,z})^{-1}}{2}$  \\            
\hline
\end{tabular}
\end{center}
\end{table*}

In order to calculate the path loss terms in Table~\ref{table_6} and~\ref{table_7} (i.e. $L_{u,x}$, $L_{v,y}$, $L_{u,z}$, $L_{v,z}$), we apply the following propagation models, specific to our deployment scenarios in Section~\ref{section: case study variants}. For the outdoor links we consider the \mbox{ITU-R} model for \mbox{line-of-sight} (LOS) propagation within street canyons and the non-line-of-sight (NLOS) model for over roof-top propagation~\cite{ITU-R2013}. For the indoor links we apply the multi-wall-and-floor (MWF) model in~\cite{LottRhode2001}, where we assume the indoor walls are 10~cm thick concrete walls (i.e. 16~dB and 14~dB attenuation through the first and the following traversed walls, respectively) and the floors are 20~cm thick concrete walls (i.e. 29~dB and 24~dB attenuation through the first and the following traversed floors, respectively). We assume the building entry loss of 19.1 dB for external walls~\cite{ITU-R2013}. For outdoor to indoor links or indoor to outdoor links we consider cascaded models of indoor and outdoor propagation models. We assume log-normal shadowing with 4~dB standard deviation for indoor links and 7~dB for all other links~\cite{3GPP2010}.

\subsection{LBT MAC Overhead}
\label{section: MAC OH}

We model the MAC overhead for \emph{LBT} due to sensing time based on the parameters of IEEE 802.11n CSMA/CA for the 5~GHz band, without RTS/CTS~\cite{IEEE2012}, by extending Bianchi's analytical model in~\cite{Bianchi2000}. 
For each AP $x$ we estimate the MAC efficiency $S_x$ in~(\ref{eq_28}) by quantifying the fraction of time the channel is used to successfully transmit frames as 
\begin{equation}\label{eq_19}
S_x=\frac{\overline{T_{f,x}}}{\overline{T_{s,x}}-\overline{T_{c,x}}+\sigma\frac{\overline{T^*_{c,x}}-(1-\tau)^n(\overline{T^*_{c,x}}-1)}{n\tau(1-\tau)^{n-1}}},
\end{equation}
where $\overline{T_{f,x}}$ is the average duration of a frame in the CS range of $x$, $\overline{T_{s,x}}$ is the average time the channel is occupied by a successful transmission in the CS range of $x$, $\overline{T_{c,x}}$ is the average time the channel is occupied by a collision in the CS range of $x$, $\sigma$=9~$\mu$s is the duration of an empty backoff time slot, $\overline{T^*_{c,x}}=\overline{T_{c,x}}/\sigma$, $n=1+|\mathbf{A}_x|+|\mathbf{B}_x|$ is the total number of APs within the CS range of AP $x$, $\tau$ is the probability that a station transmits in a randomly chosen time slot. We calculate a lookup table for $\tau$ for each value of $n$ based on Bianchi's model for binary exponential backoff with $CW_{min}$=15 and $CW_{max}$=1023~\cite{IEEE2012}.
The terms $\overline{T_{f,x}}$, $\overline{T_{s,x}}$, and $\overline{T_{c,x}}$ are defined analogously as   
\begin{equation}\label{eq_20}
\overline{T_{f/s/c,x}}=\frac{T_{f/s/c,x}+ \displaystyle \sum_{z\in\mathbf{A}_x}T_{f/s/c,z} + \sum_{z\in\mathbf{B}_x}T_{f/s/c,z}}{1+|\mathbf{A}_x|+|\mathbf{B}_x|},
\end{equation}
where $z$ represents other APs in $x$'s CS range.

For an AP $x$, we define the duration of a frame $T_{f,x}$, the time the channel is kept busy due to successful transmission $T_{s,x}$, and the time AP $x$ occupies the channel due to a collision $T_{c,x}$ as
\begin{equation}\label{eq_23}
T_{f,x}=\begin{cases}
   \begin{split} PHY&_{header}+\frac{MAC_{header}+MSDU}{R_x}, \\ &\text{if $x$ has \emph{LBT} and IEEE 802.11n PHY} \end{split}\\
   \begin{split} 1 \text{ ms}&, \\&\text{if $x$ has \emph{LBT} and LTE PHY} \end{split} 
   \end{cases}
\end{equation}
\begin{equation}\label{eq_24}
T_{s,x}=\begin{cases}
    \begin{split} T_{f,x}&+DIFS+SIFS+PHY_{header}+\frac{ACK}{R_{WiFi,min}}, \\&\text{if $x$ has \emph{LBT} and IEEE 802.11n PHY}  \end{split}\\
    \begin{split} T_{f,x}&+DIFS, \\&\text{if $x$ has \emph{LBT} and LTE PHY}  \end{split}
    \end{cases}
\end{equation}
\begin{equation}\label{eq_25}
T_{c,x}=\begin{cases}
    \begin{split} T_{f,x}&+DIFS, \\&\text{if $x$ is has \emph{LBT} and IEEE 802.11n PHY} \end{split}\\
    \begin{split} T_{f,x}&+DIFS, \\&\text{if $x$ is has \emph{LBT} and LTE PHY,} \end{split}
    \end{cases}
\end{equation}
where $R_x$ is the transmission rate of AP $x$, $R_{WiFi,min}$=6.5~Mbps, $SIFS$=16~$\mu$s, $DIFS=SIFS+2\times\sigma$=34~$\mu$s, $ACK$=112~bits, $PHY_{header}$=40~$\mu$s, $MAC_{header}$=112~bits (including FCS)~\cite{IEEE2012}, and $MSDU$=1500~Bytes~\cite{Alliance2015}. 
We note that the frame duration $T_{f,x}$ of an AP $x$ with \emph{LBT} and LTE PHY is fixed to 1~ms (i.e. the duration of an LTE subframe). 

\subsection{Throughput Degradation due to Frame Collisions when Implementing Duty Cycle}
\label{section: col model}
We assume a worst case scenario, where all \emph{LBT} frames from a population $\mathbf{M}$ transmitted at the end of a \emph{duty cycle} time slot collide with \emph{duty cycle} frames from a population $\mathbf{N}$ and are lost, if there is a \emph{duty cycle} frame transmitted in the next time slot within the CS range. For our case study we assume only \emph{LBT} frames are lost, since the APs implementing \emph{duty cycle} have a better spectral efficiency and their users are able to decode frames at a lower SINR. 

The throughput degradation of AP $x$ in $\mathbf{M}$, $r^{M,N}_{deg}(x)$ in Table~\ref{table_4}, is given by 
\begin{equation}\label{eq_26}
r^{M,N}_{deg}(x)=\begin{cases}
	\begin{split} \frac{1}{m},~\text{if $\mathbf{N}$ has \emph{50\% duty cycle} } \end{split}\\
	\begin{split} \frac{1}{m} \times &\Bigg [1-\displaystyle\prod_{z\in{\mathbf{B}_x}} \bigg(1-\frac{1}{|\mathbf{C}_z|+|\mathbf{D}_z|}\bigg) \Bigg], \\&\text{if $\mathbf{N}$ has \emph{adaptive duty cycle,}} \end{split}
\end{cases}
\end{equation}
where $m$ is the total number of \emph{LBT} frames of $\mathbf{M}$ that can be transmitted in one \emph{duty cycle} time slot and the other parameters are defined in Table~\ref{table_1}. The duration of an \emph{LBT} frame is typically lower than a time slot (i.e. the \emph{duty cycle} ON-duration), so $m>1$. For a median IEEE 802.11n rate of 32.5~Mbps, the duration of a complete \emph{LBT} frame transmission is $T_{s,x}$=419~$\mu$s. Assuming a time slot duration\footnote{Our simulation results showed that the throughput degradation does not visibly vary with the considered duty cycle \mbox{ON-duration}, therefore in Section~\ref{section: results} we will only present the results for the 100~ms variant.}
 of either 10~ms (i.e. the duration of an LTE frame) or 100 ms (i.e. maximum ON-duration specified by Qualcomm~\cite{Qualcomm2015}), then $m=23$ or $m=238$. 

If the APs in $\mathbf{N}$ implement \emph{fixed 50\% duty cycle}, every time slot containing a transmission of $\mathbf{M}$ is followed by a time slot containing a transmission of $\mathbf{N}$, therefore at the end of a time slot containing transmissions of $\mathbf{M}$ a frame collision will always occur. Within a time slot there are multiple \emph{LBT} frames of $\mathbf{M}$ transmitted and each AP in $\mathbf{M}$ within the CS range transmits a roughly equal number of frames per time slot.    

If the APs in $\mathbf{N}$ implement \emph{adaptive duty cycle}, it is not necessary that a time slot containing transmissions of $\mathbf{M}$ is followed by a time slot containing transmissions of $\mathbf{N}$, since the \emph{adaptive duty cycle} transmissions may overlap in time, possibly leaving more consecutive time slots unoccupied. The term $\Bigg [1-\displaystyle\prod_{z\in{\mathbf{B}_x}} \bigg(1-\frac{1}{|\mathbf{C}_z|+|\mathbf{D}_z|}\bigg) \Bigg]$ is the probability that the next time slot contains a \emph{duty cycle} transmission of $\mathbf{N}$.

Throughput degradation due to collisions between frames of APs in $\mathbf{M}$ and $\mathbf{N}$ implementing \emph{ideal TDMA} is considered negligible, as the transmissions of $\mathbf{N}$ can be scheduled such that the alternation of slots containing frames of $\mathbf{M}$ and frames of $\mathbf{N}$ is reduced.

 
\section{Results and Analysis of Coexistence Case Study}
\label{section: results}

We conduct extensive Monte Carlo simulations in Matlab with 3000 network realizations for the \emph{indoor/indoor} scenarios \emph{with} or \emph{without internal walls}, and 1500 network realizations for the \emph{indoor/outdoor} and \emph{outdoor/outdoor} scenarios.
We evaluate the throughput performance of the two populations of APs in terms of the downlink throughput\footnote{We consider the downlink as this is relevant for the current work on LTE in unlicensed to transmit only downlink user data traffic. We also note that throughput, aside from being the fundamental network performance evaluation metric \emph{in general}, is also considered as the primary performance metric in major LTE/Wi-Fi coexistence studies, e.g.~\cite{Forum2015, Alliance2015}.} per AP, based on our throughput and interference model in Section~\ref{section: interf}.  
We present a representative selection of our results with respect to the key parameters in our framework in Section~\ref{section: mechanisms}, in order to derive \emph{general} insights into \mbox{inter-technology} coexistence in a spectrum commons.
Firstly, we consider the effect of channel selection schemes and the extent of interference coupling given by the considered deployment scenarios. 
Since the MAC is arguably the most important component of spectrum sharing mechanisms, we then focus on analysing in detail the considered \emph{LBT} and \emph{duty cycle} MAC mechanisms, with respect to the identified cases of interference coupling. Finally, we consider the effect of the PHY capabilities.

\subsection{Impact of Channel Selection Schemes: Sense and Random}
\label{ch sel}

We compare the \emph{sense} and \emph{random} channel selection schemes as described in Section~\ref{section: case study variants}, in order to study the extent to which they mitigate inter-technology interference. 
\begin{figure}[t!]
\centering
\subfigure[Median throughput per legacy AP.]
          {\includegraphics[width=0.9\linewidth]{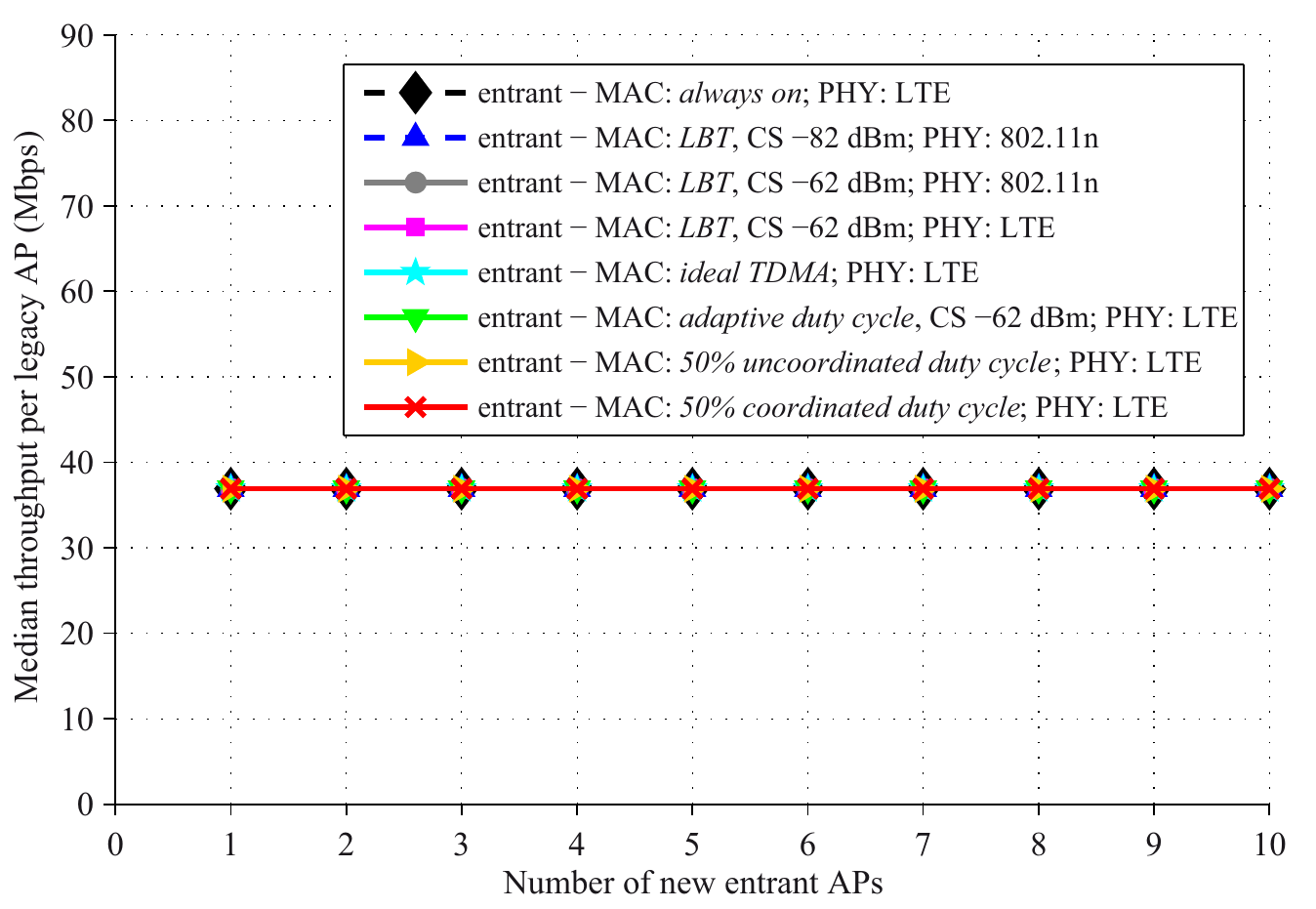}
           \label{fig_1_1}}
\\
\subfigure[Median throughput per entrant AP.]      	
	   	{\includegraphics[width=0.9\linewidth]{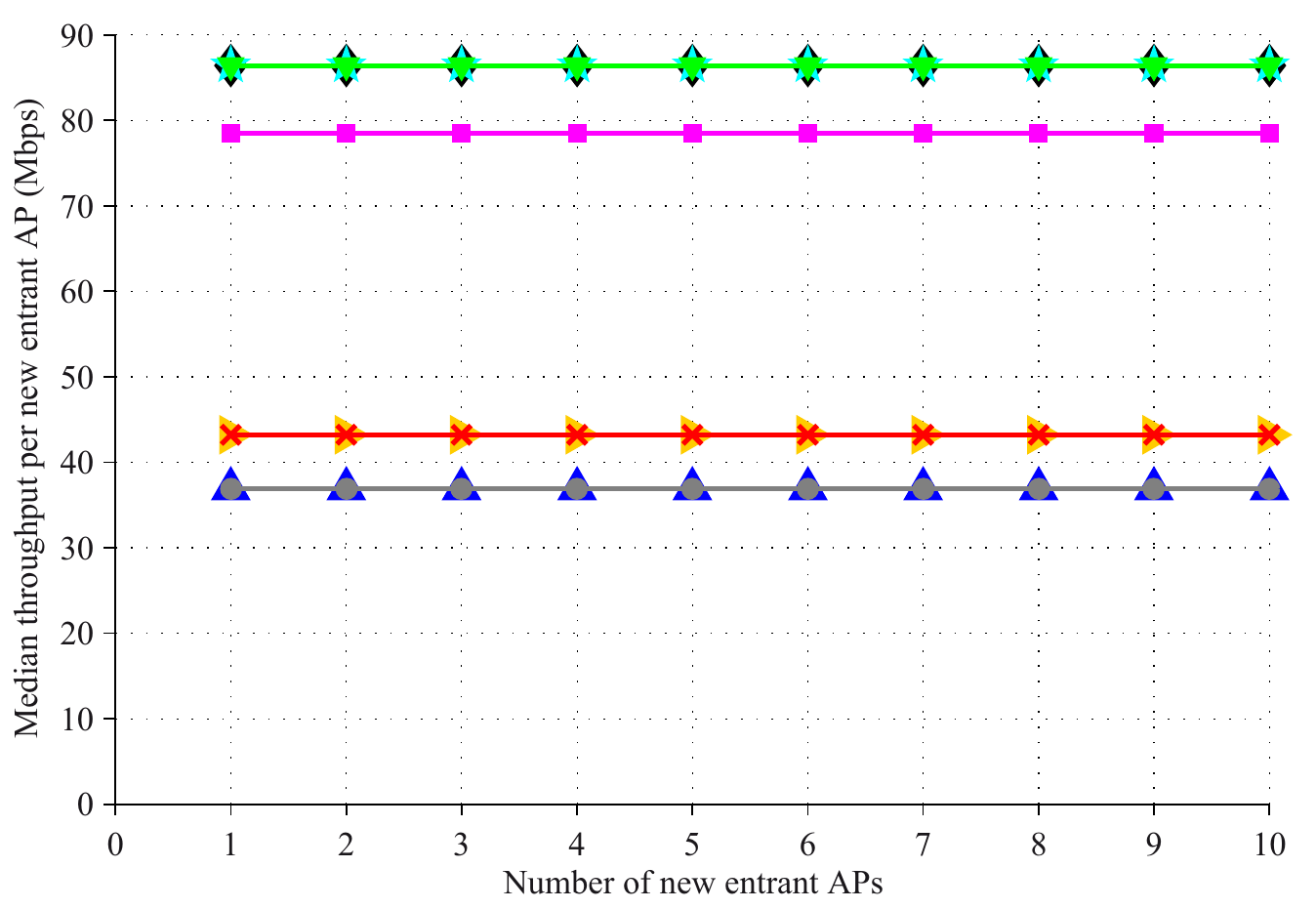}
		   \label{fig_1_2}}
\caption{Median throughput per legacy and new entrant AP with different entrant spectrum sharing mechanisms, for the \emph{indoor/indoor} scenario with \emph{sense}, for 10 legacy and 1-10 entrant APs.}
\label{fig_1}
\end{figure}
Fig.~\ref{fig_1} shows the median throughput, over all Monte Carlo realizations, of legacy and new entrant APs for the \emph{indoor/indoor} scenario with \emph{sense}, for 10 legacy APs and a variable number of entrant APs. 
The legacy AP throughput is constant at 36.9~Mbps, irrespective of the entrant AP density or spectrum sharing mechanism, due to the high number of available channels in the 5~GHz band. Consequently, there is a low number of APs per channel, so the spectrum \mbox{time-sharing} mechanisms are not triggered at all. Similarly, the entrant AP throughput is not affected by the AP density, but is instead simply determined by the entrant's own implemented MAC and PHY layer. More specifically, \emph{always on}, \emph{ideal TDMA}, and \emph{adaptive duty cycle} with LTE PHY achieve the maximum throughput of 86.4~Mbps, whereas the throughput for \emph{LBT} with LTE PHY is slightly lower (78.4~Mbps), due to the \emph{LBT} sensing overhead, $S_x$. Both variants of \emph{fixed 50\% duty cycle} achieve half of the maximum LTE PHY throughput, as expected. \emph{LBT} with IEEE 802.11n PHY achieves the lowest throughput (36.9~Mbps) of the considered entrant technologies, due to its lower PHY spectral efficiency.

\begin{figure}[t!]
\centering
\subfigure[Throughput distribution for legacy APs.]
          {\includegraphics[width=0.9\linewidth]{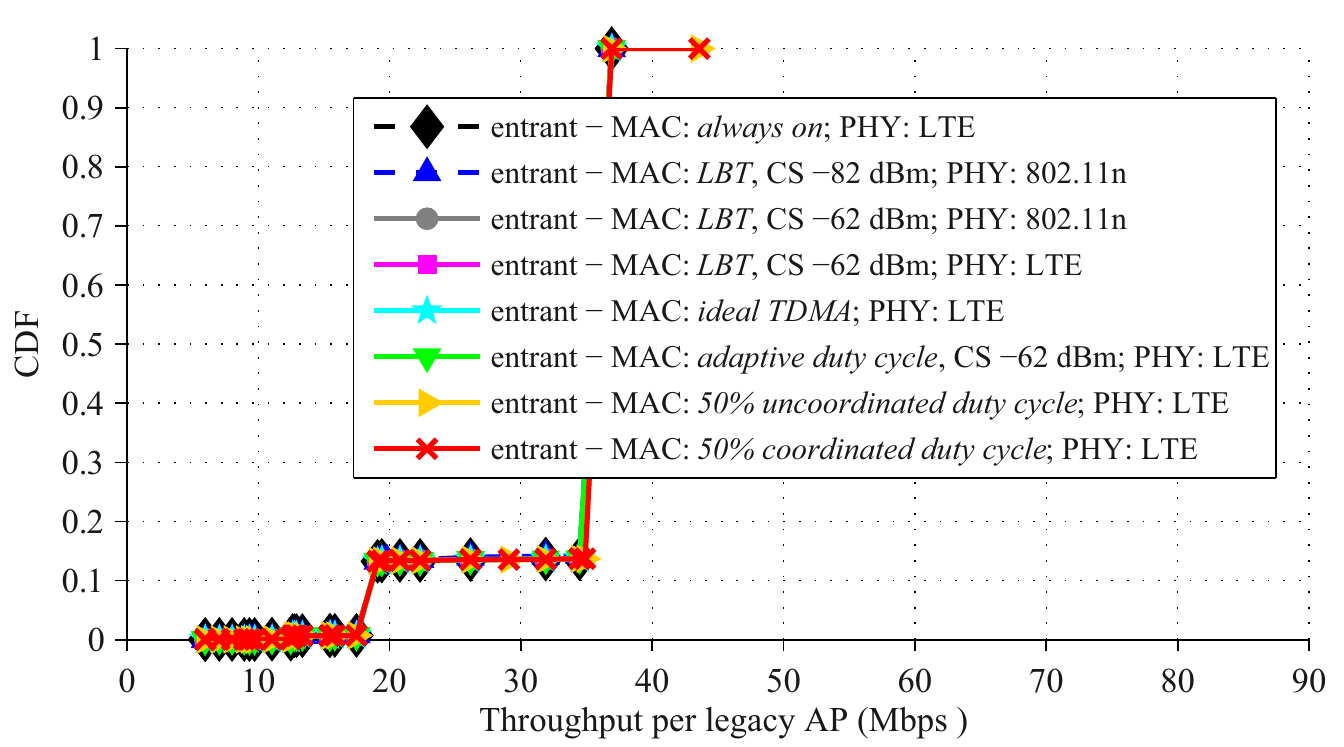}
           \label{fig_cdf_1}}
\\
\subfigure[Throughput distribution for entrant APs.]      	
	   	{\includegraphics[width=0.9\linewidth]{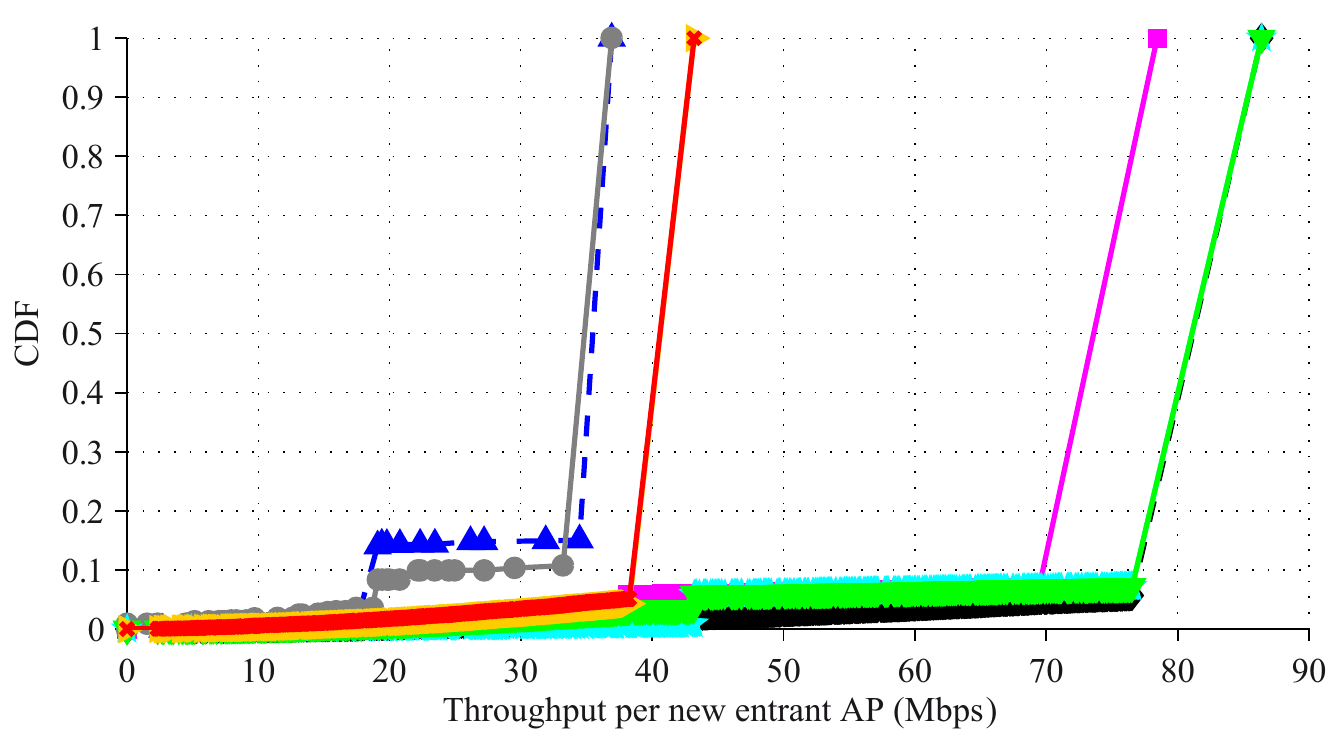}
		   \label{fig_cdf_2}}
\caption{Throughput distribution for legacy and new entrant APs with different entrant spectrum sharing mechanisms, for the \emph{indoor/indoor} scenario with \emph{sense}, for 10 legacy and 10 new entrant APs.}
\label{fig_cdf}
\end{figure}

Let us now consider in more detail the most challenging coexistence case (i.e. 10 legacy and 10 new entrant APs). 
Fig.~\ref{fig_cdf} shows the throughput distribution over all Monte Carlo realizations, for the \emph{indoor/indoor} scenario, for \emph{sense}. 
The distribution of the legacy AP throughput in Fig.~\ref{fig_cdf_1} is the same for all entrant MAC schemes, which shows that \mbox{co-channel} transmissions are always avoided for legacy and entrant APs, for our considered realistic network densities. 
Fig.~\ref{fig_cdf_2} shows that no more than 15\% of the entrant APs share the channel with another entrant. Otherwise, the throughput is identical to the median throughput in Fig.~\ref{fig_1_2}.
Namely, the performance of the spectrum sharing mechanisms indicated by the median throughput per AP is largely representative of trends over the whole throughput distribution\footnote{We note the median throughput per AP was consistently found representative of the overall distribution in our results, so in the remainder of this paper we will focus on the median throughput only, such that we can consider the effect of increasing entrant density, as in Fig.~\ref{fig_1}.}.

The throughput results for \emph{sense} are qualitatively similar for all other considered scenarios and AP densities. 
The results for \emph{random} and \emph{sense} are also in general comparable, except for few cases with strong co-channel interference, where \emph{sense} eliminates the interference between the two populations of APs (e.g. the \emph{indoor/indoor} scenario \emph{without internal walls}). We omit explicitly presenting these results, for the sake of brevity.

Given the high number of available channels in the 5~GHz band, interference is largely avoided by simply allocating different channels to different APs. It follows that, in practice, LTE and \mbox{Wi-Fi} could harmoniously coexist in the 5~GHz band, regardless of the MAC mechanism implemented by LTE. In the remaining sections, we will thus only focus on the \emph{single channel} results, in order to study in general the performance of spectrum \mbox{time-sharing} techniques.  

\subsection{Impact of the Extent of Network-Wide Interference Coupling between APs}
\label{interf coupling}

\begin{figure*}[t!]
\centering
\subfigure[Median throughput per legacy AP for \emph{indoor/indoor}.]
          {\includegraphics[width=0.45\linewidth]{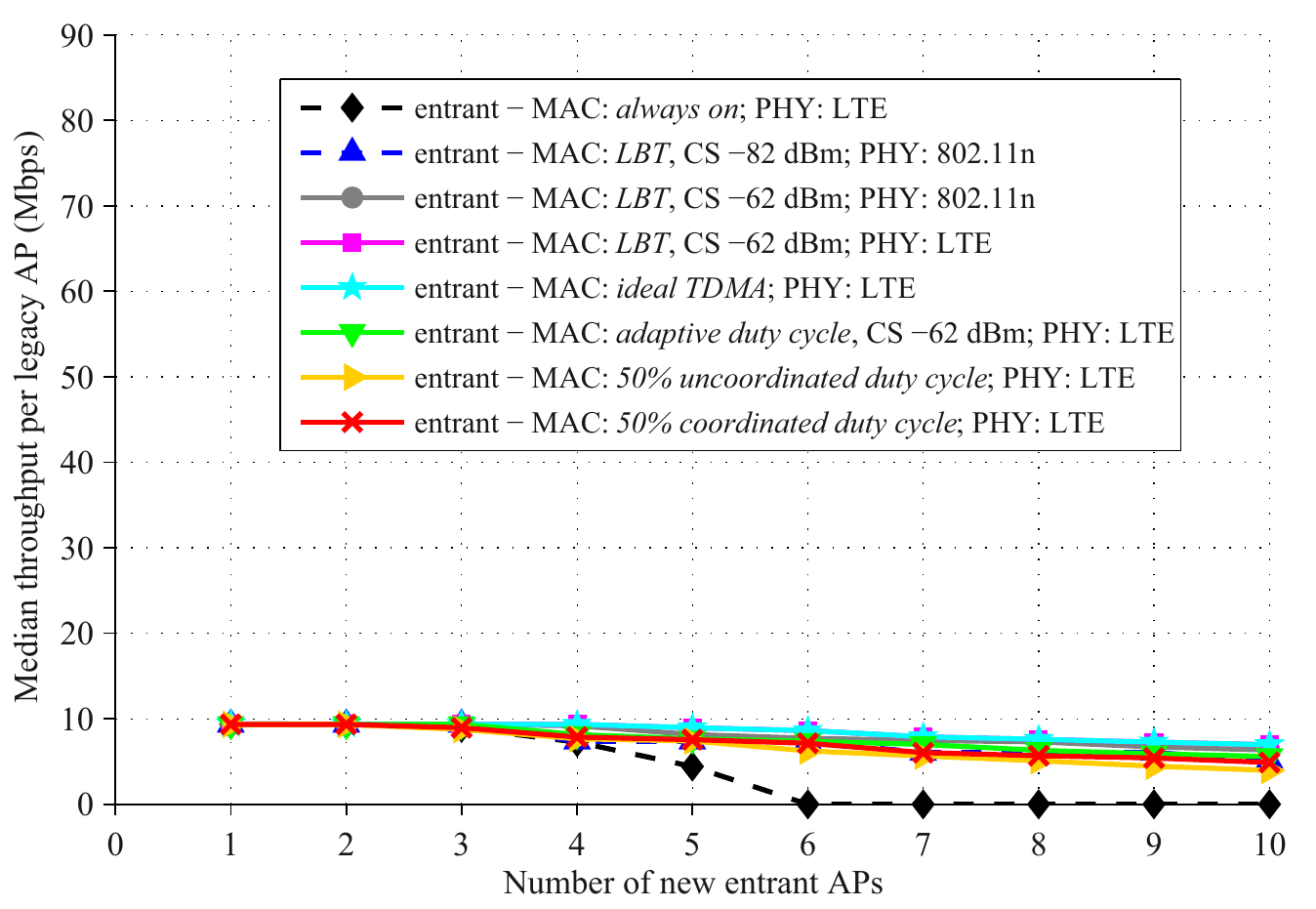}
           \label{fig_2_1}}
~
\subfigure[Median throughput per entrant AP for \emph{indoor/indoor}.]   
	  	   {\includegraphics[width=0.45\linewidth]{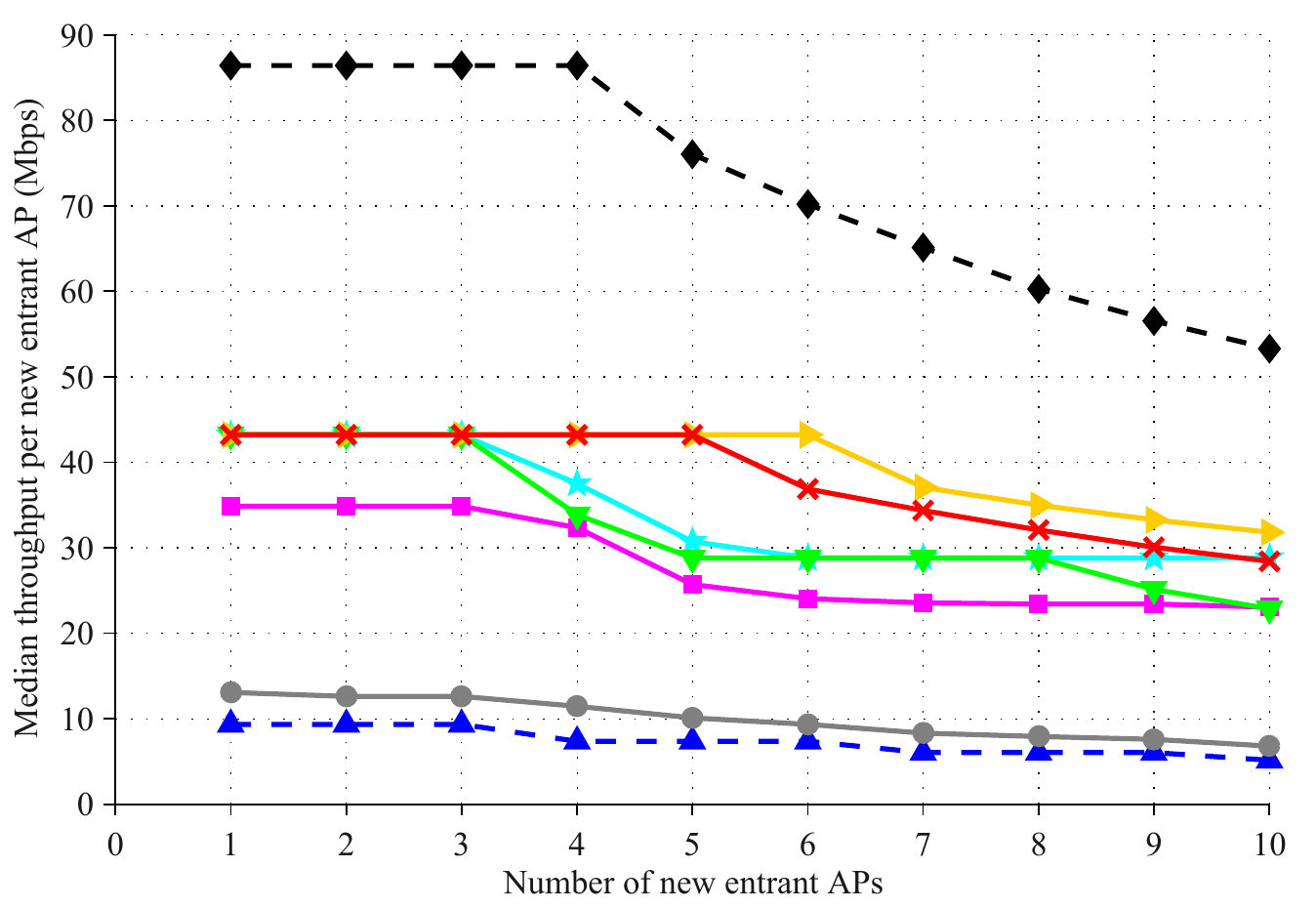}
		   \label{fig_2_2}}
\\
\subfigure[Median throughput per legacy AP for \emph{indoor/indoor without internal walls}.]
          {\includegraphics[width=0.45\linewidth]{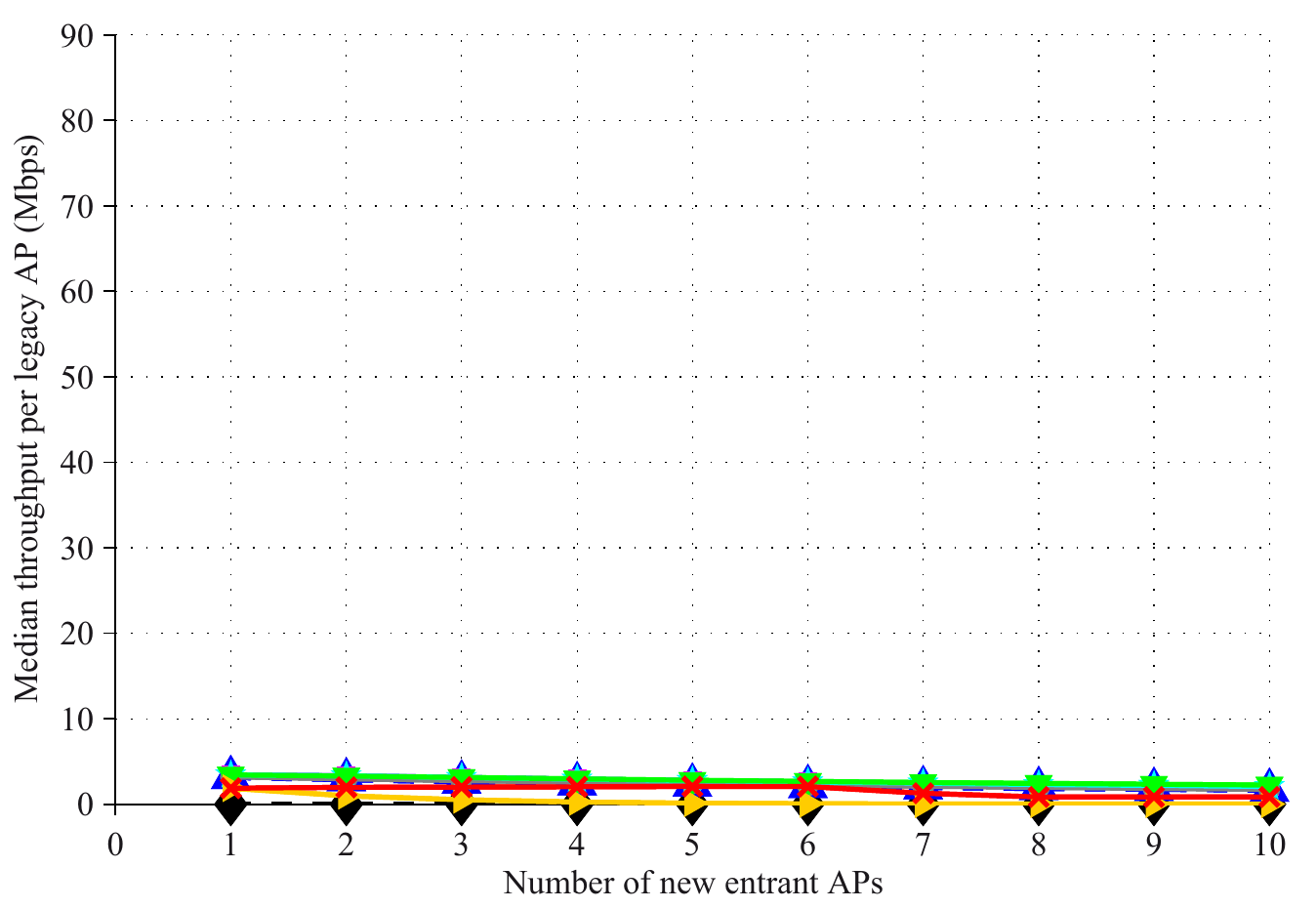}
           \label{fig_2_3}}
~
\subfigure[Median throughput per entrant AP for \emph{indoor/indoor without internal walls}.]      	  	   		       		 
          {\includegraphics[width=0.45\linewidth]{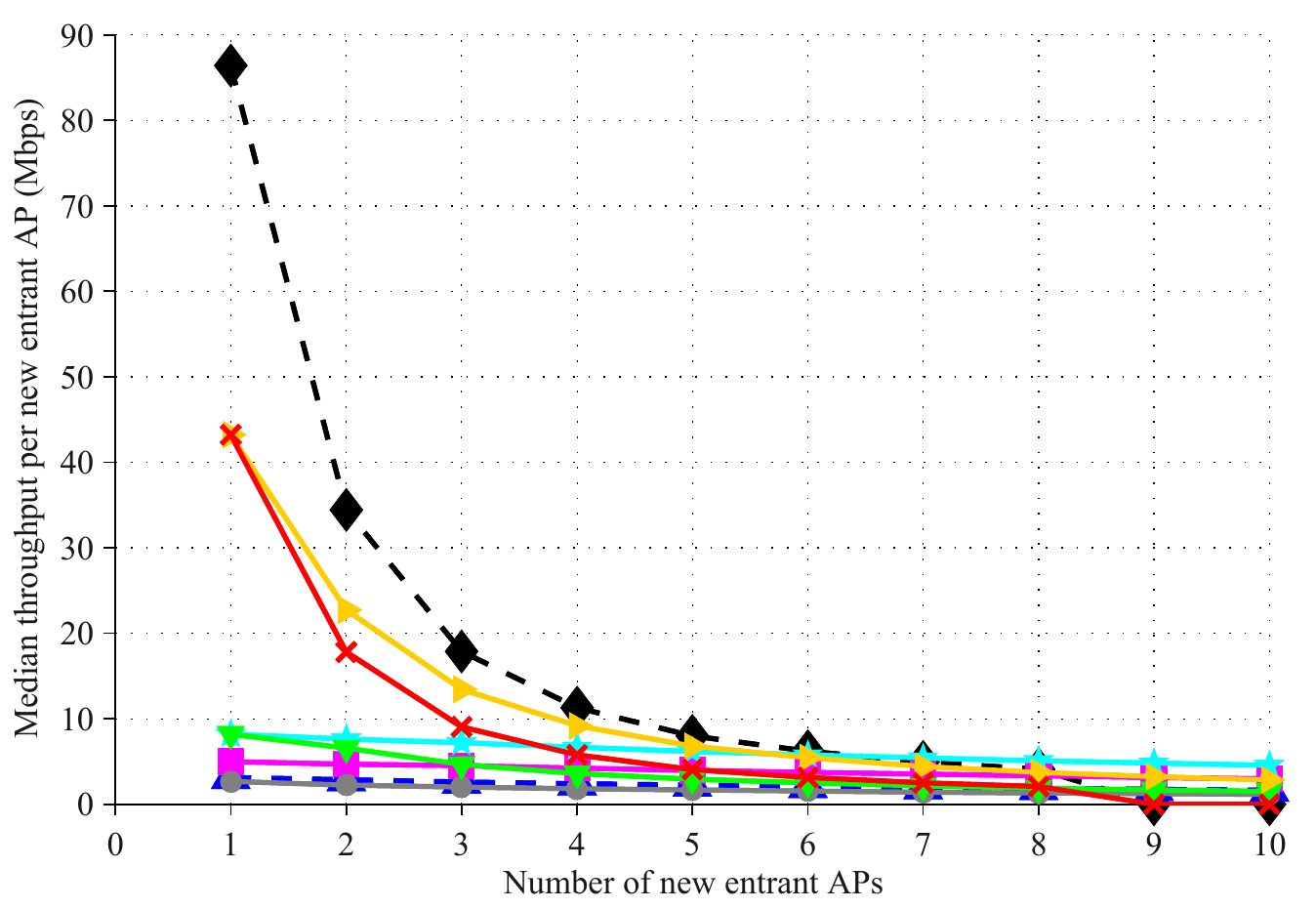}
		   \label{fig_2_4}}
\caption{Median throughput per legacy and new entrant AP with different entrant spectrum sharing mechanisms, for the \emph{indoor/indoor} scenario \emph{with} and \emph{without internal walls} with \emph{single channel}, for 10 legacy and 1-10 entrant APs.}
\label{fig_2}
\end{figure*} 

In this section, we study the effect of wall shielding, as discussed in Section~\ref{section: case study variants}, on the interference coupling among APs.
Fig.~\ref{fig_2} shows the median throughput per legacy and entrant AP over all network realizations for the \emph{indoor/indoor} scenarios \emph{with} and \emph{without internal walls}, with \emph{single channel}, for 10 legacy and 1-10 entrant APs. Comparing Figs.~\ref{fig_2_1} and \ref{fig_2_3} and Figs.~\ref{fig_2_2} and \ref{fig_2_4}, the shielding from the internal walls results in a throughput increase of up to 10~Mbps and 75~Mbps for the legacy and new entrant APs, respectively, when the entrants do not implement any spectrum \mbox{time-sharing} mechanism (i.e. \emph{always on}). Additionally, a major and consistent improvement is shown for all other spectrum sharing mechanisms, for high building shielding, since in such cases the number of APs within CS range is reduced, so that coexistence must be managed between fewer APs. 
We emphasize that the legacy AP throughput is considerably degraded (down to 0~Mbps) when coexisting with entrants with \emph{always on}, even for high wall shielding conditions in the \emph{indoor/indoor} scenario in Fig.~\ref{fig_2_1}. This indicates that a \mbox{time-sharing} MAC mechanism should always be imposed for coexisting devices.  

\begin{figure*}[t!]
\centering
\subfigure[Median throughput per legacy AP, for 5000 legacy APs/km\textsuperscript{2} and 2-20 entrant APs, for \emph{indoor/outdoor}.]
          {\includegraphics[width=0.45\linewidth]{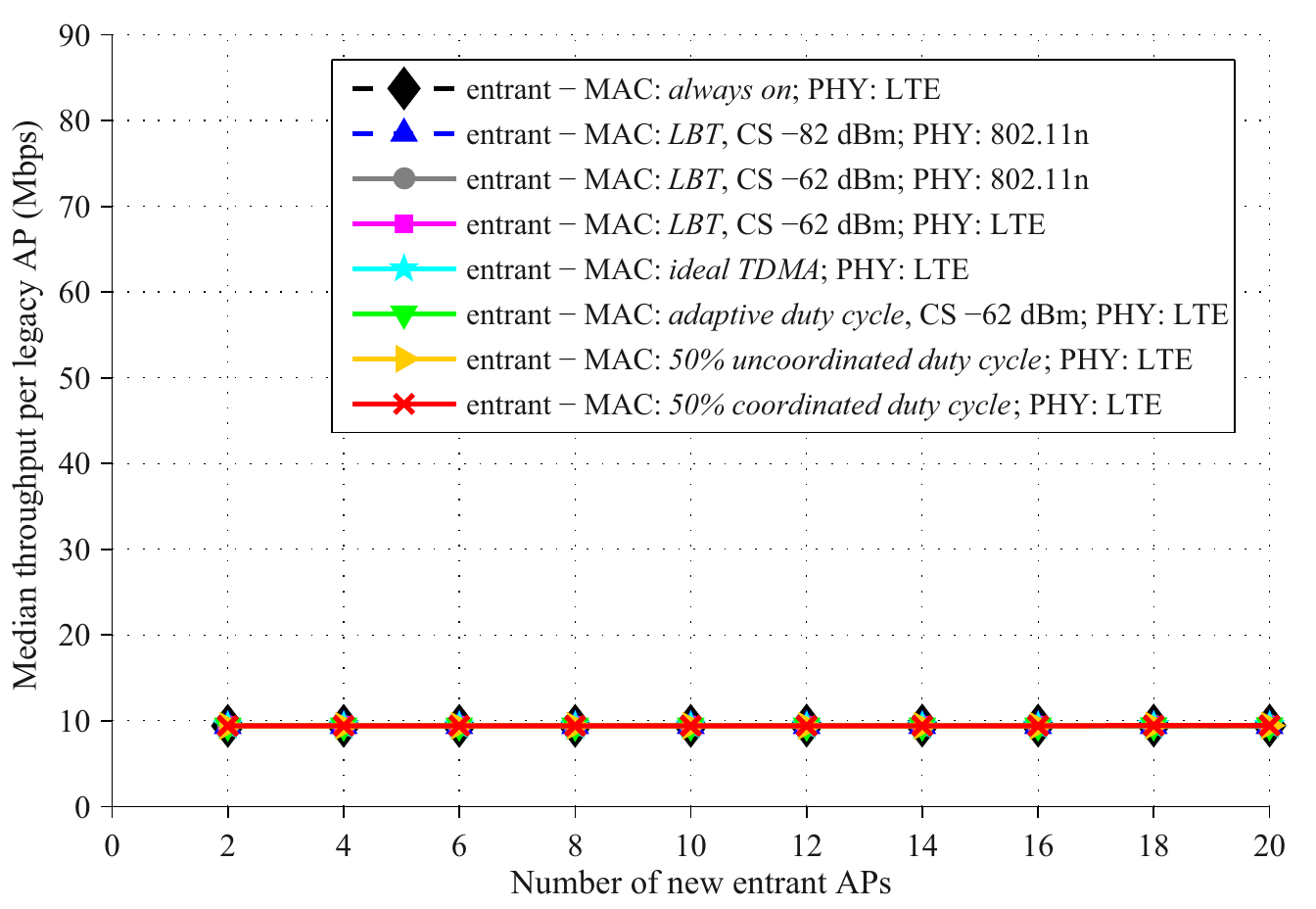}
           \label{fig_5_1}}
~
\subfigure[Median throughput per entrant AP, for 5000 legacy APs/km\textsuperscript{2} and 2-20 entrant APs, for \emph{indoor/outdoor}.]      	  	  {\includegraphics[width=0.45\linewidth]{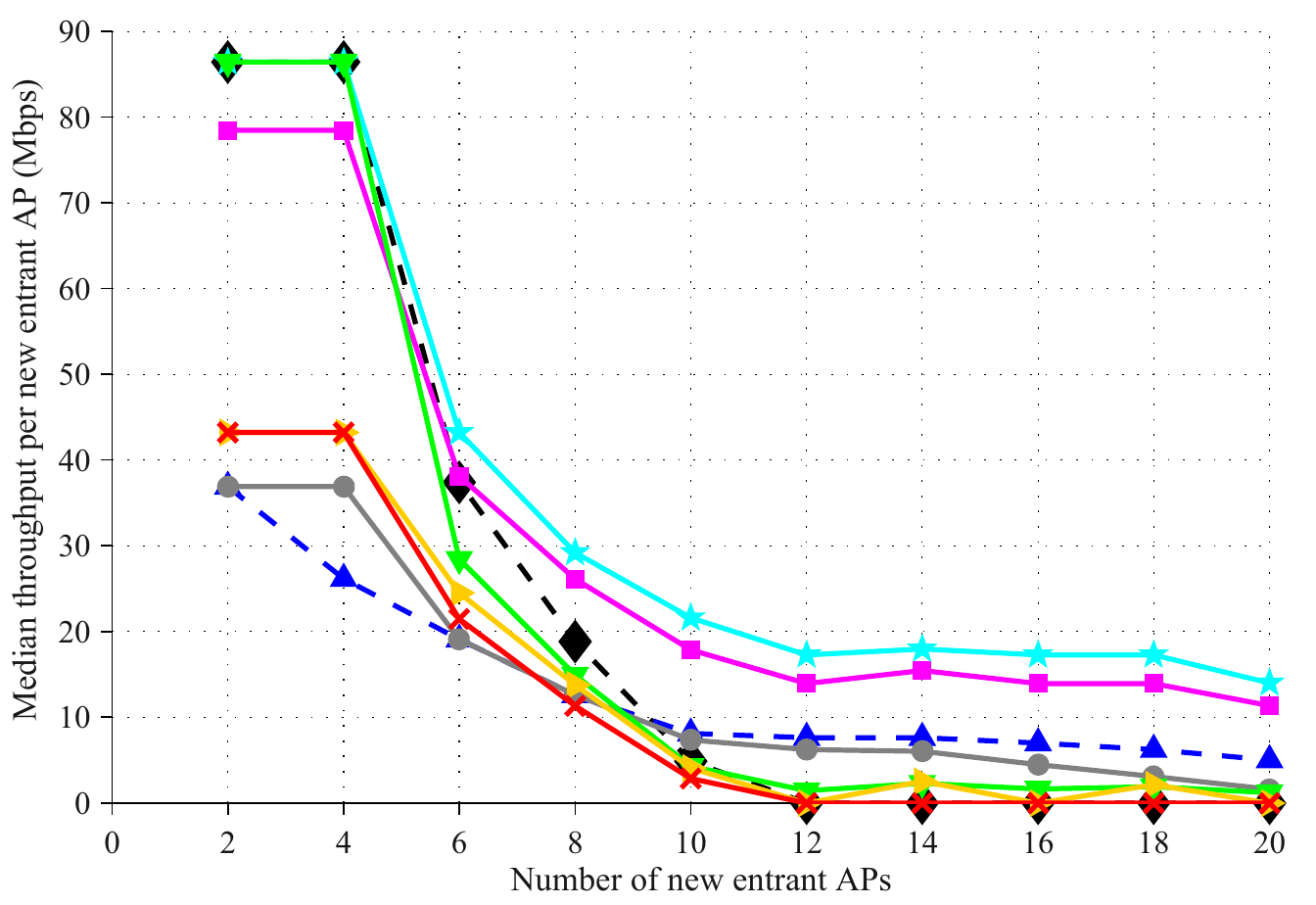}
		   \label{fig_5_2}}
\\
\subfigure[Median throughput per legacy AP, for 10 legacy APs and 1-10 entrant APs, for \emph{outdoor/outdoor}.]
          {\includegraphics[width=0.45\linewidth]{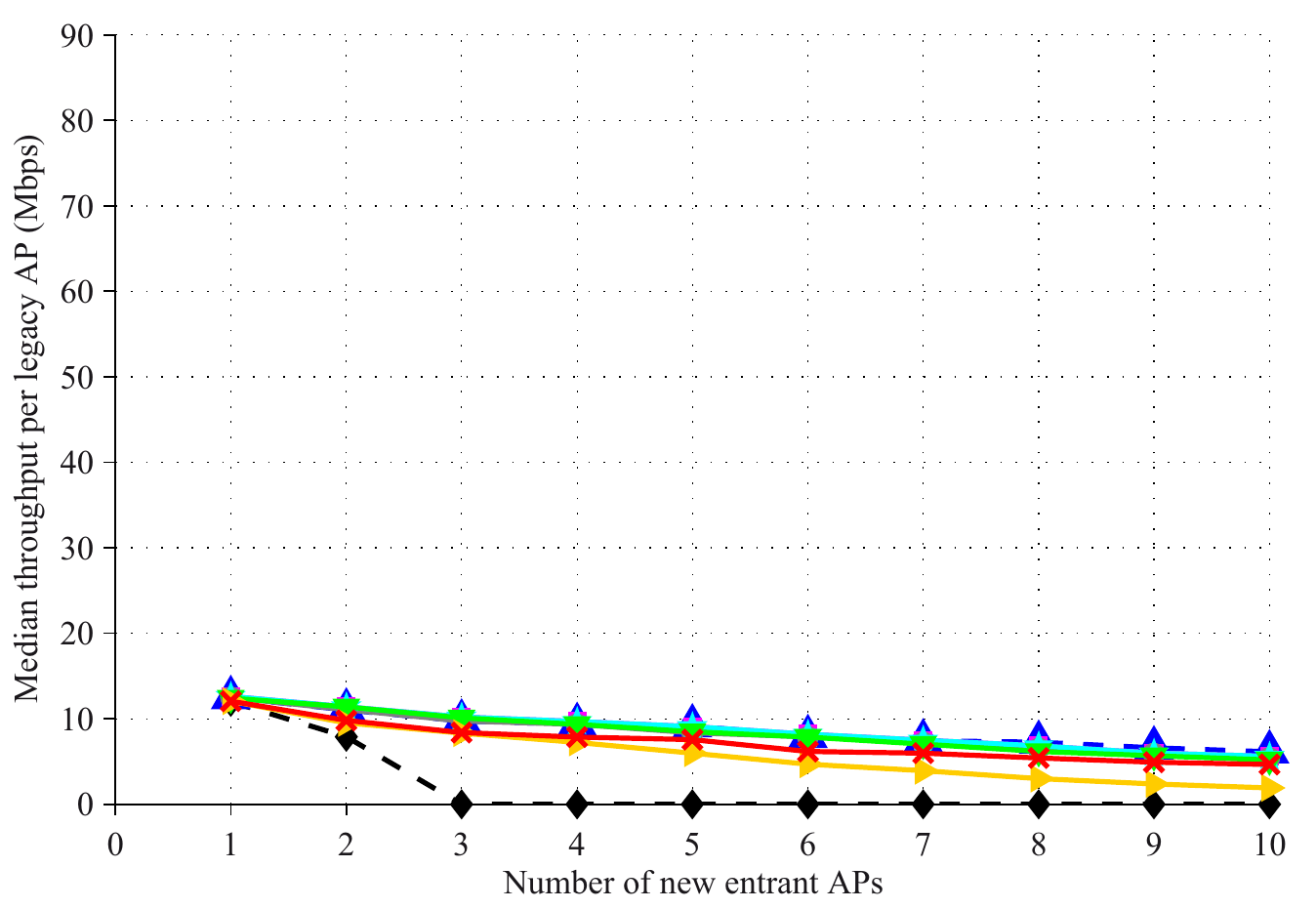}
           \label{fig_5_3}}
~
\subfigure[Median throughput per entrant AP, for 10 legacy APs and 1-10 entrant APs, for \emph{outdoor/outdoor}.]      	 
        {\includegraphics[width=0.45\linewidth]{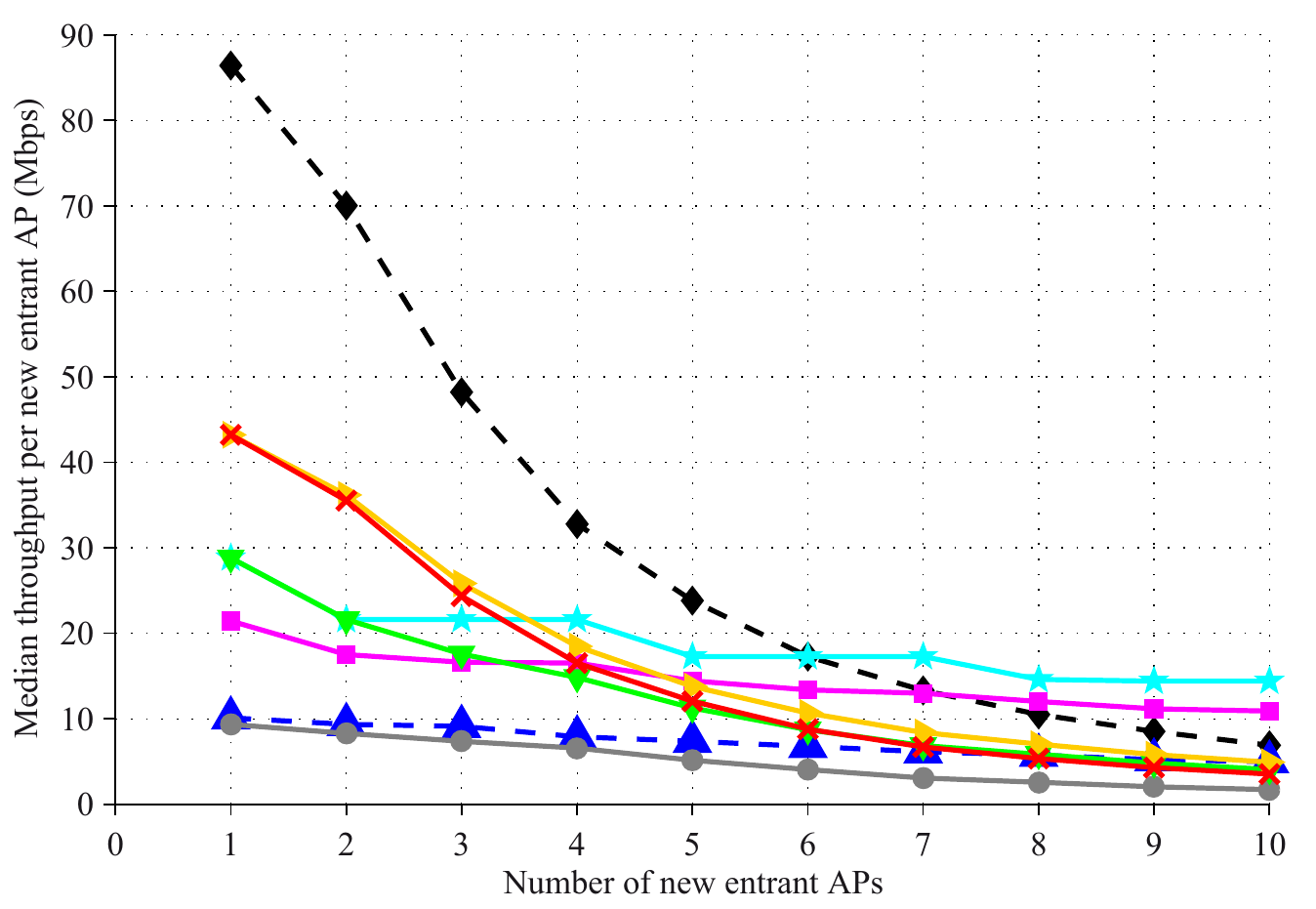}
		   \label{fig_5_4}}
\caption{Median throughput per legacy and entrant AP with different entrant spectrum sharing mechanisms, for the \emph{indoor/outdoor} and \emph{outdoor/outdoor} scenarios, for \emph{single channel}.}
\label{fig_5}
\end{figure*}

Fig.~\ref{fig_5} shows the median throughput per legacy and new entrant AP over all network realizations for the \emph{indoor/outdoor} and \emph{outdoor/outdoor} scenarios with \emph{single channel}. The legacy AP throughput for the \emph{indoor/outdoor} scenario in Fig.~\ref{fig_5_1} is invariant with the entrant AP density at 10~Mbps. Consistent with these results, we have observed the entrant AP throughput to be similar for both 5000 and 500 legacy APs/km\textsuperscript{2} (results not shown here for brevity), demonstrating that the indoor and outdoor APs are isolated from each other~\cite{VoicuLondon2015}. 
For the \emph{outdoor/outdoor} scenario in Fig.~\ref{fig_5_3}, the legacy AP throughput decreases by up to 10~Mbps compared to the \emph{indoor/outdoor} scenario in Fig.~\ref{fig_5_1}. Also, the entrant AP throughput for the \emph{outdoor/outdoor} scenario in Fig.~\ref{fig_5_4} decreases by up to 16~Mbps compared to the \emph{indoor/outdoor} scenario in Fig.~\ref{fig_5_2}, for the same number of entrant APs, due to the low shielding between the two populations of APs when both coexist outdoors. These results suggest that, in practice, coexistence can easily be ensured in indoor residential scenarios, or between APs in indoor deployments and outdoor hotspot deployments, due to the presence of high building shielding. Conversely, within \mbox{open-plan} indoor hotspot scenarios, or outdoor hotspot scenarios, the MAC mechanisms should be carefully selected, in order to improve the coexistence performance.  
We continue discussion of Figs.~\ref{fig_2} and~\ref{fig_5} in Sections~\ref{results MAC}--\ref{PHY}.

\subsection{Impact of Fundamental Choice of MAC Scheme: LBT vs. Adaptive Duty Cycle}
\label{results MAC}

In this section we compare the performance of two different MAC approaches, \emph{LBT} and \emph{adaptive duty cycle}, which nevertheless both aim to achieve equal share of air time for APs located within CS range. However, the major difference between these schemes is that \emph{LBT} guarantees an interference-free CS range at the expense of additional MAC overhead, whereas \emph{adaptive duty cycle} eliminates overhead due to sensing time, but cannot avoid interference in the CS range in a non-cooperative manner. We analyse how this design tradeoff for the two schemes affects the coexistence performance.   

Fig.~\ref{fig_2_2} shows that the entrant AP throughput for \emph{adaptive duty cycle} is consistently around 5~Mbps higher than for \emph{LBT} with the same CS threshold of \mbox{-62}~dBm and LTE PHY. 
Therefore, for higher shielding between APs in the \emph{indoor/indoor} scenario, and thus a lower number of APs in the CS range, \emph{adaptive duty cycle} consistently achieves a higher throughput than \emph{LBT}.
For scenarios with lower shielding between legacy and entrant, or among entrant APs, Figs.~\ref{fig_2_4}, \ref{fig_5_2}, and \ref{fig_5_4} show a \emph{switching point} at a critical number of entrant APs, after which the median throughput obtained for \emph{LBT} becomes higher than for \emph{adaptive duty cycle}.
This demonstrates that for low interference coupling, the \emph{LBT} sensing overhead degrades the throughput more than the actual interference does, especially when this interference is reduced by \emph{adaptive duty cycle}, due to its inherent mechanism of randomly selecting a time slot to transmit in. By contrast, for high interference coupling scenarios, the interference experienced by a given AP becomes too strong to be efficiently managed by \emph{adaptive duty cycle} and thus avoiding it within the CS range at the expense of additional \emph{LBT} sensing overhead becomes more beneficial.

\begin{figure}[t!]
\centering
\subfigure[Median throughput per legacy AP.]
          {\includegraphics[width=0.9\linewidth]{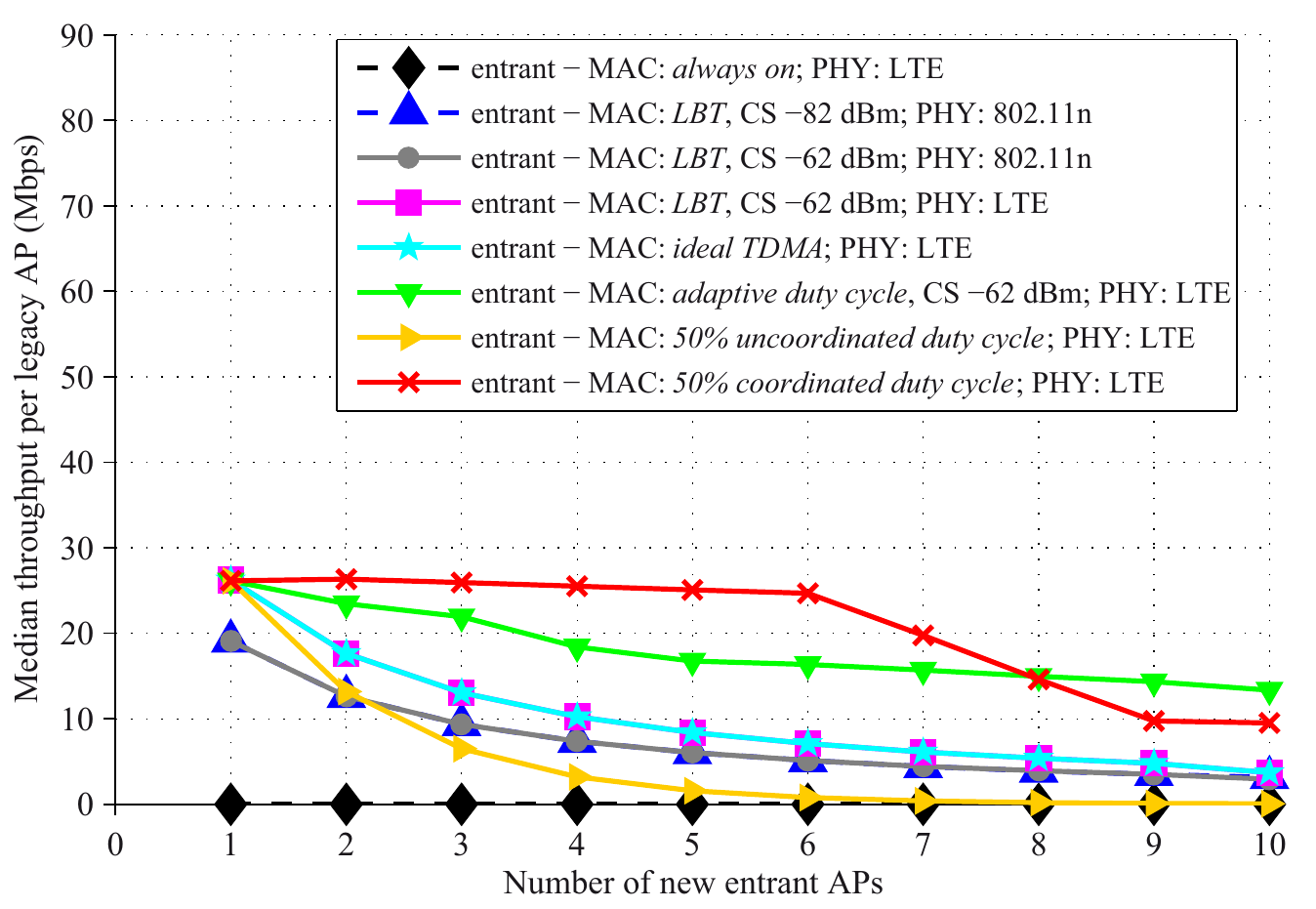}
           \label{fig_4_1}}
\\
\subfigure[Median throughput per entrant AP.]
	   	{\includegraphics[width=0.9\linewidth]{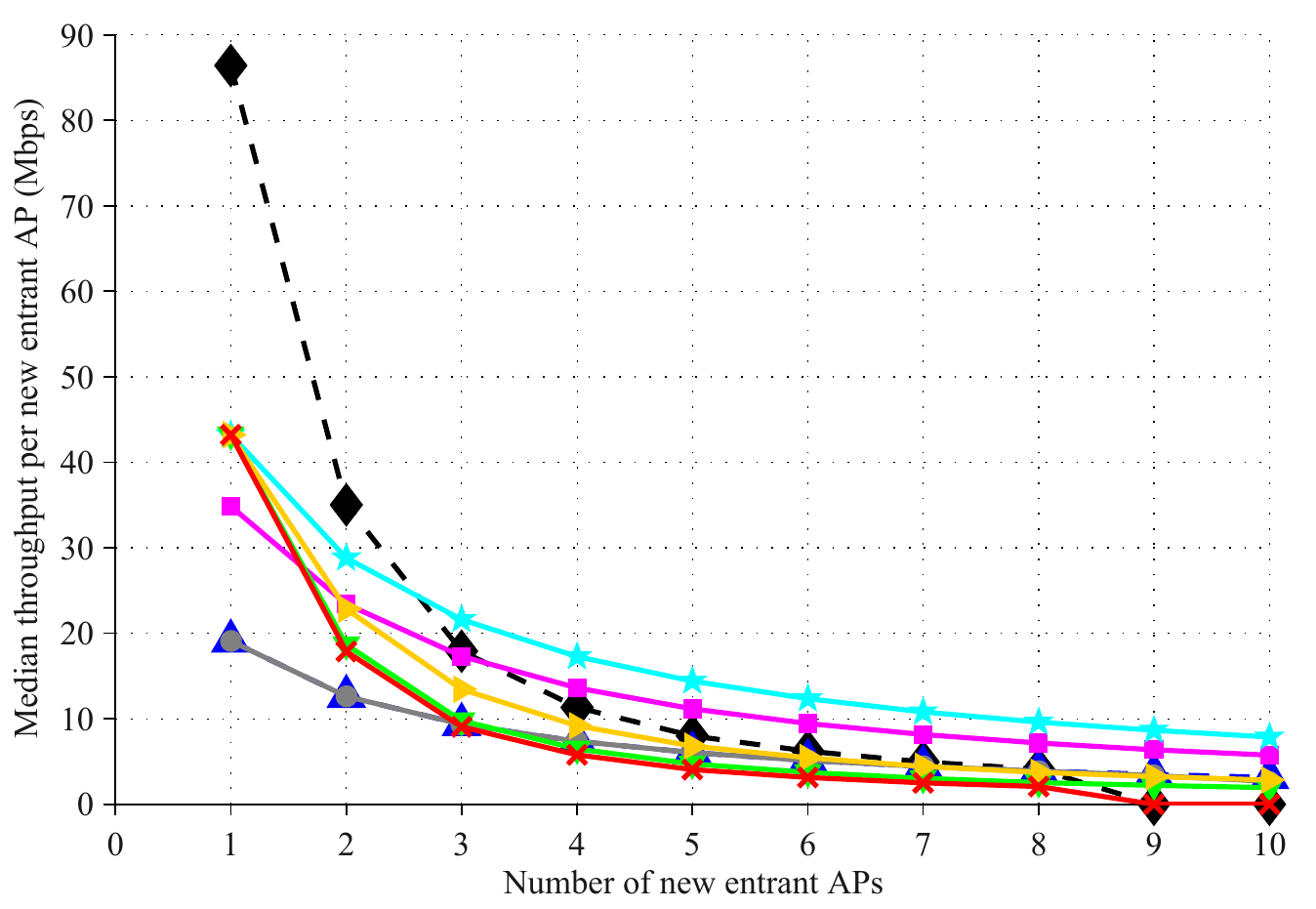}
		   \label{fig_4_2}}
\caption{Median throughput per legacy and entrant AP with different entrant spectrum sharing mechanisms, for the \emph{indoor/indoor} scenario \emph{without internal walls} with \emph{single channel}, for 1 legacy and 1-10 entrant APs.}
\label{fig_4}
\end{figure} 

Figs.~\ref{fig_2_1}, \ref{fig_2_3}, \ref{fig_5_1}, and \ref{fig_5_3} show that the legacy AP throughput does not vary significantly if the entrants implement \emph{LBT} or \emph{adaptive duty cycle}, due to the fact that the legacy APs implement \emph{LBT} and will thus not experience interference from within the CS range in either case. 
However, for the border case of one legacy AP in the \emph{indoor/indoor} scenario \emph{without internal walls}, Fig.~\ref{fig_4_1} shows an increase of up to 10~Mbps in the median legacy AP throughput when the entrants implement \emph{adaptive duty cycle} compared to \emph{LBT}. As the entrant APs randomly transmit when implementing \emph{adaptive duty cycle}, their transmissions may overlap in time, leaving more air time for the legacy AP implementing \emph{LBT}. 
This tradeoff can also be seen for the entrants in Fig.~\ref{fig_4_2}, where the \emph{adaptive duty cycle} median throughput drops quickly below the \emph{LBT} throughput for only 2 entrant APs. Nevertheless, this increase in legacy AP throughput is minor, due to the presence of other legacy APs with which the additional air time is shared.   

Our results thus demonstrate that for high interference coupling, \emph{LBT} outperforms \emph{adaptive duty cycle} due to its capability of better protecting both its own AP and other coexisting technologies against interference. Conversely, for low interference coupling, \emph{adaptive duty cycle} outperforms \emph{LBT}, since it does not incurr additional MAC sensing overhead.     

\subsection{Impact of LBT Parameters: CS Threshold, Frame Duration Type, and MAC Overhead}
\label{LBT}

In this section we study the effect of the \emph{LBT} CS threshold, frame duration types, and sensing overhead for \emph{LBT}, as given in Section~\ref{section: MAC OH} on the coexistence performance.    

Fig.~\ref{fig_2_2} shows that in the \emph{indoor/indoor} scenario a higher CS threshold of -62~dBm yields a higher entrant AP throughput than -82~dBm, by comparing the two \emph{LBT} variants with IEEE 802.11n PHY. The corresponding throughput difference is almost constant at about 2~Mbps when increasing the number of APs.   
For lower shielding between entrant APs, such as in the \emph{indoor/outdoor} scenario in Fig.~\ref{fig_5_2} and the \emph{outdoor/outdoor} scenario in Fig.~\ref{fig_5_4}, there is a switching point at a critical number of APs, after which the throughput for the -82~dBm CS threshold is higher than for -62~dBm. 
These results show how the CS threshold controls the tradeoff between sharing the channel in time within the CS range and suffering from interference from outside the CS range. For low interference coupling, the APs implementing a low CS threshold unnecessarily defer to other APs. However, in case of strong interference, a lower CS threshold protects the users better.
Our results thus indicate that it is beneficial to adapt the CS threshold according to the individually experience interference per AP, since it consistently affects the throughput performance (albeit not by a large margin).

Figs.~\ref{fig_2_1}, \ref{fig_2_3}, \ref{fig_5_1}, and \ref{fig_5_3} show that the legacy AP throughput is the same when the entrants transmit frames of different duration types, as seen by comparing the throughput for \emph{LBT} with -62~dBm and LTE PHY (i.e. fixed frame duration) against \emph{LBT} with -62~dBm and IEEE 802.11n PHY (i.e. rate-based frame duration). Namely, the difference in the MAC overhead term $S_x$ (\emph{cf.} Section~\ref{section: MAC OH}) is marginal for different frame duration types. 
A difference of at most 7~Mbps is evident for the border case of 10 entrants coexisting with only 1 legacy AP in Fig.~\ref{fig_4_1}. Since the fixed frame duration is longer than the rate-based duration, the sensing time is shorter relative to the transmission time, resulting in a slightly higher $S_x$ within the CS range.  
The typical frame duration is thus not an important parameter for the coexistence performance.  

Our results for the entrant APs in Figs.~\ref{fig_2_2}, \ref{fig_2_4}, \ref{fig_5_2}, and \ref{fig_5_4} show that the \emph{LBT} MAC overhead is not significantly high, as evident by comparing against \emph{ideal TDMA}, which is a perfectly coordinated time-sharing MAC without sensing overhead or interference within the CS range. 
The throughput per entrant AP implementing \emph{ideal TDMA} is higher than for \emph{LBT} by up to about 10~Mbps (for high wall shielding in Fig.~\ref{fig_2_2}), corresponding to the \emph{LBT} MAC overhead. Although this difference occurs consistently, it becomes negligible (down to about 2~Mbps) for \mbox{low-shielding} dense networks in Figs.~\ref{fig_2_4}, \ref{fig_5_2}, and \ref{fig_5_4}.  
This indicates that the distributed \emph{LBT} scheme performs almost as well as an ideal \mbox{time-sharing} scheme for high AP densities, so that, in such cases, intra- and \mbox{inter-operator} coordination is not worthwhile. However, for low AP densities, such coordination becomes beneficial.    

\subsection{Impact of Duty Cycle Parameters: Adaptiveness to AP Detection and Coordination Level}
\label{duty cycle}

In this section we evaluate the effect of key \emph{duty cycle} MAC parameters on the coexistence performance: the ability to adapt the duty cycle when detecting other APs (i.e. \emph{fixed 50\%} vs. \emph{adaptive duty cycle}), and the level of local coordination (i.e. \emph{uncoordinated} vs. \emph{coordinated} \emph{fixed 50\% duty cycle}, and \emph{adaptive duty cycle} vs. \emph{ideal TDMA}).     

For the \emph{indoor/indoor} scenario in Fig.~\ref{fig_2_2} and the \emph{outdoor/outdoor} scenario in Fig.~\ref{fig_5_4}, the entrant AP throughput for \emph{fixed 50\% duty cycle} is higher than that for \emph{adaptive duty cycle} by up to 15~Mbps. The contrary holds for the \emph{indoor/outdoor} scenario in Fig.~\ref{fig_5_2}, where \emph{adaptive duty cycle} outperforms \emph{fixed 50\% duty cycle} by up to 30~Mbps. This difference in trend occurs because in the \emph{indoor/outdoor} scenario the entrant APs coexist only among themselves, due to the isolation given by the wall shielding between the two populations of APs. If \emph{fixed 50\% duty cycle} APs coexist among themselves for high network densities, they will also experience increased interference compared to \emph{adaptive duty cycle} APs coexisting among themselves, due to the increased likelihood of having overlapping transmissions. Instead, in the \emph{indoor/indoor} and \emph{outdoor/outdoor} scenarios, the entrant APs with \emph{fixed 50\% duty cycle} also coexist with legacy APs implementing \emph{LBT}, a mechanism which avoids interference within the CS range. In such cases, \emph{fixed 50\% duty cycle} APs are protected against interference and also have more and a fixed number of transmission opportunities (i.e. half of the time) compared to \emph{adaptive duty cycle}, which tries to protect coexisting \emph{LBT} APs. This also shows that \emph{LBT} protects its own APs, as well as other coexisting technologies. 
Regardless of the coexisting legacy population, it becomes evident that \emph{adaptive duty cycle} outperforms \emph{fixed 50\% duty cycle} in case of very high network densities (i.e. increased interference coupling) because it reduces the interference for other entrant APs within the CS range. For very low network densities in Fig.~\ref{fig_5_2}, \emph{adaptive duty cycle} also outperforms \emph{fixed 50\% duty cycle}. Otherwise, for moderate interference coupling, \emph{fixed 50\% duty cycle} can suffer from interference and still achieve better throughput results than \emph{adaptive duty cycle}, due to its constant air time.
This indicates that adapting the \emph{duty cycle} to the number of APs within CS range is important for both the APs implementing this scheme and other coexisting \emph{LBT} APs, for the entire range of considered network densities. 

Our results consistently show in Figs.~\ref{fig_2_2}, \ref{fig_2_4}, \ref{fig_5_2}, and \ref{fig_5_4} that for the entrant throughput there is only a marginal difference between \emph{fixed 50\% coordinated duty cycle} and \emph{fixed 50\% uncoordinated duty cycle}. For legacy APs, coexistence with \emph{fixed 50\% uncoordinated duty cycle} can become problematic in scenarios like \emph{indoor/indoor without internal walls} in Fig.~\ref{fig_4_1}, where the legacy AP throughput for the \emph{uncoordinated} variant drops down to 0~Mbps, whereas the throughput for the \emph{coordinated} variant is up to 25~Mbps higher. Therefore, coordinating \emph{fixed 50\% duty cycle} APs such that they transmit at the same time, may compensate for this MAC scheme's lack of adaptiveness in coexistence scenarios.

Finally, Figs.~\ref{fig_2_2} and \ref{fig_2_4} show that increasing the coordination level for \emph{adaptive duty cycle} (i.e. \emph{ideal TDMA}) does not increase the entrant AP throughput significantly in the \emph{indoor/indoor} scenario \emph{with} or \emph{without internal walls} (at most 5~Mbps).   
However, in the \emph{indoor/outdoor} scenario in Fig.~\ref{fig_5_2} and \emph{outdoor/outdoor} scenarios in Fig.~\ref{fig_5_4}, the entrant AP throughput is by up to 15~Mbps higher for \emph{ideal TDMA} than for \emph{adaptive duty cycle}. This is important especially in high network density cases where the entrant throughput for \emph{adaptive duty cycle} drops down to 0~Mbps.   
These results demonstrate that intra- and \mbox{inter-operator} coordination of \emph{adaptive duty cycle} would bring no significant benefits in case of low interference coupling, but are worthwhile in case of high interference coupling. However, intra- and inter-operator coordination would not always be possible in  practice, and would also increase the control management overhead, so that distributed spectrum sharing may still be more attractive.

\subsection{Impact of PHY Spectral Efficiency}
\label{PHY}

Lastly, we consider the effect of the PHY-MAC interactions on the coexistence performance. 
Figs.~\ref{fig_2_2}, \ref{fig_2_4}, \ref{fig_5_2}, and \ref{fig_5_4} show a consistent and significant difference in terms of entrant AP throughput, of up to 40~Mbps, between \emph{LBT} with LTE PHY and \emph{LBT} with IEEE 802.11n PHY. These results are simply due to the better spectral efficiency of LTE PHY vs. IEEE 802.11n PHY. 
Fig.~\ref{fig_5_4} in particular shows that a CS threshold of \mbox{-82}~dBm achieves a higher throughput than -62~dBm for IEEE 802.11n PHY (i.e. -82~dBm is preferred), but that -62~dBm with LTE PHY largely outperforms -82~dBm with IEEE 802.11n PHY, regardless of its poorly performing CS threshold. This demonstrates that the superior LTE PHY can in fact compensate for a \mbox{sub-optimal} CS threshold of \emph{LBT}.
Although we vary the PHY layer only for \emph{LBT} MAC variants, we would expect to obtain similar qualitative results also for other MAC mechanisms. 
Our results suggest that a high-performing PHY layer may not only have a direct, substantial, and consistent impact on the throughput performance, but can also compensate for MAC parameters that are loosely tuned.

\section{Conclusions}
\label{section: conclusion}

In this paper we presented a detailed, systematic, and transparent study of different distributed spectrum sharing mechanisms for \mbox{inter-technology} coexistence in a spectrum commons. 
Firstly, we proposed a general framework for comparatively evaluating these spectrum sharing mechanisms, by identifying the key constituent design parameters and investigating their individual effect. 
Secondly, we proposed a novel unified \mbox{network-level} throughput and interference model that captures these key design parameters per device. 
Finally, we presented a coexistence case study of two dominant technologies in a spectrum commons, i.e. \mbox{Wi-Fi} and LTE in the 5~GHz unlicensed band. 
Our extensive Monte Carlo simulation results show that LTE/\mbox{Wi-Fi} coexistence can be easily ensured though channel selection schemes, such that \mbox{time-sharing} MAC mechanisms are irrelevant. 
Moreover, our analysis of the case study results is generalizable and can be extended to other \mbox{inter-technology} coexistence cases. 
We show that, in general, the coexistence performance of MAC sharing mechanisms strongly depends on the \emph{interference coupling}, predominantly determined by building shielding, thereby identifying two regimes: (i)~low interference coupling, e.g. residential indoor scenarios, where \emph{adaptive duty cycle} outperforms \emph{LBT}, as it does not suffer from additional sensing overhead; and (ii) high interference coupling, e.g. open-plan indoor or outdoor hotspot scenarios, where \emph{LBT} outperforms \emph{adaptive duty cycle}, as it avoids strong interference within the CS range. 
We also show that, although applying intra- and \mbox{inter-operator} coordination is worthwhile in high interference coupling scenarios, the resulting gains over \emph{LBT} are minor.
Therefore, distributed MAC schemes may remain more attractive in practice.
Our ongoing work focuses on extending our study to gain further insight into time-domain parameters, including different sensing durations and backoff schemes.


%



\section*{Acknowledgment}

The authors would like to thank Dr Pierre de Vries and Prof. Petri M\"ah\"onen for useful discussions and to acknowledge funding from Deutsche Forschungsgemeinschaft (DFG).

\ifCLASSOPTIONcaptionsoff
  \newpage
\fi



\bibliographystyle{IEEEtran}
\bibliography{IEEEabrv,bibliography}
%
%
%
%

%
\begin{IEEEbiography}[{\includegraphics[width=1in,height=1.25in,clip,keepaspectratio]{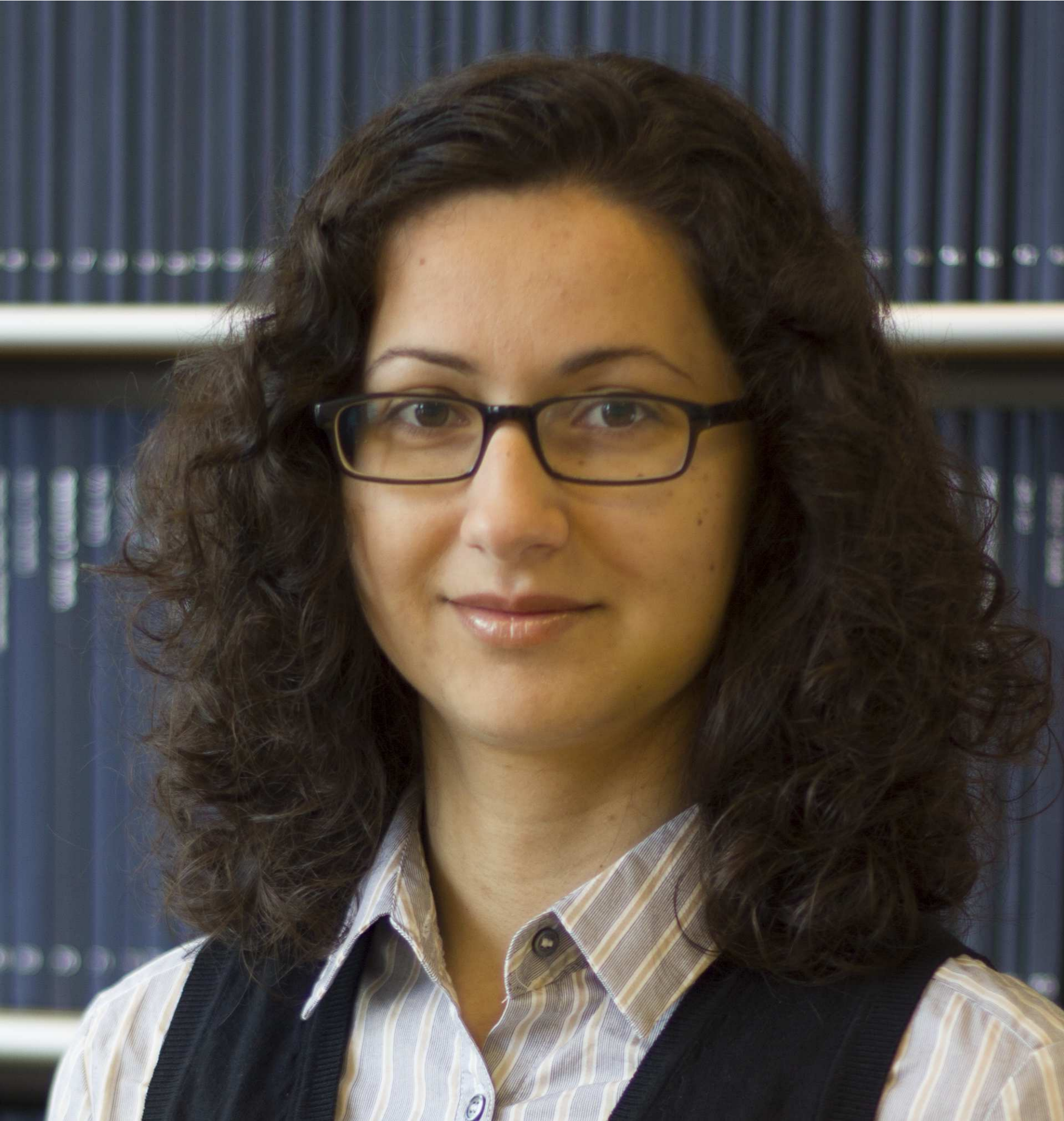}}]{Andra M. Voicu}
received her B.Sc. in electronics and telecommunications from the University Politehnica of Bucharest in 2011 and her M.Sc. in communications engineering from the RWTH Aachen University in 2013. She is currently a Ph.D. student at the Institute for Networked Systems, RWTH Aachen University, and her research work focuses on small-cell networks and distributed wireless networks.
\end{IEEEbiography}

\begin{IEEEbiography}[{\includegraphics[width=1in,height=1.25in,clip,keepaspectratio]{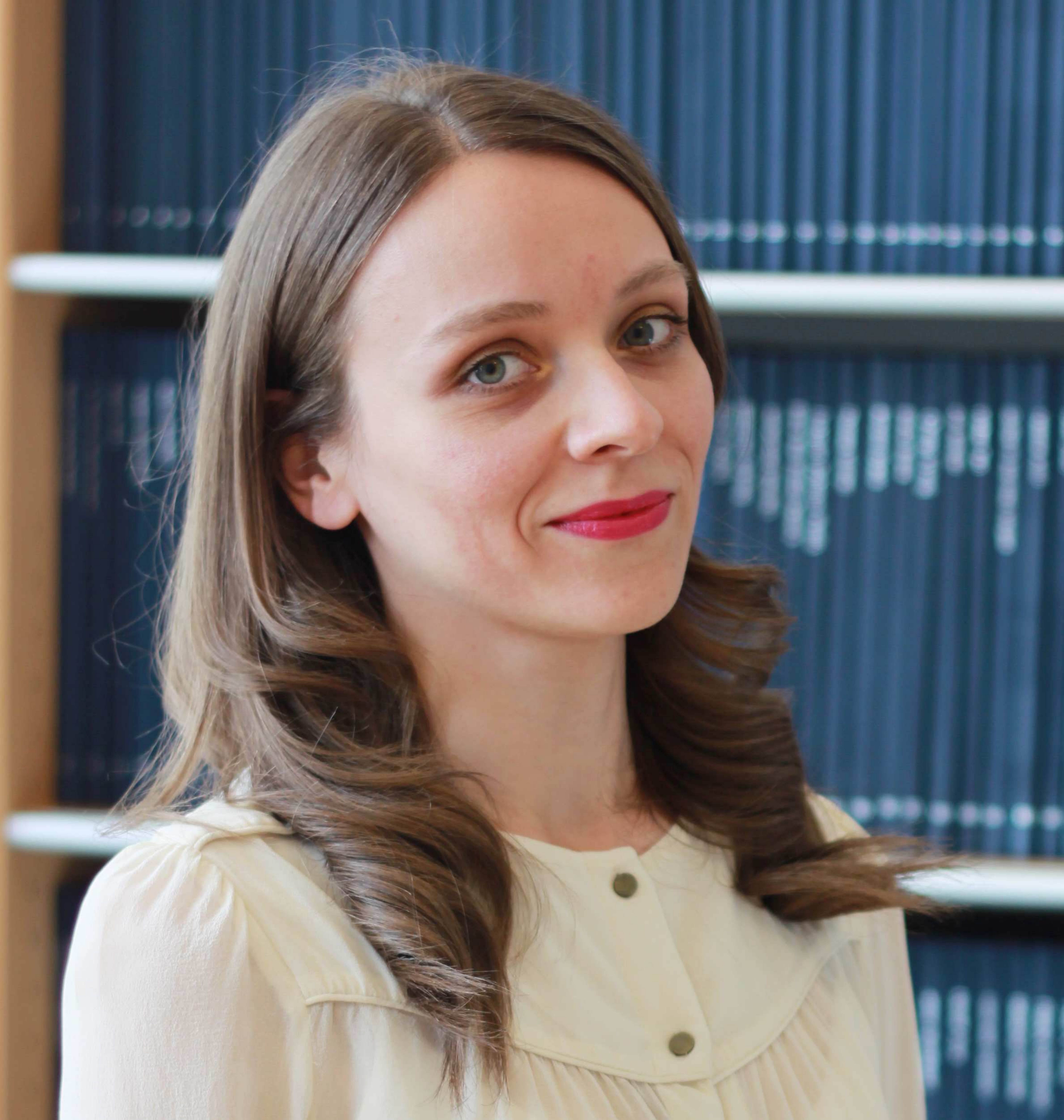}}]{Ljiljana Simi\'c}
is currently working as research coordinator and senior researcher at the Institute for Networked Systems at RWTH Aachen University. She received her Bachelor of Engineering (with 1st Class Honours) and Doctor of Philosophy degrees in Electrical and Electronic Engineering from The University of Auckland in 2006 and 2011, respectively. Prior to joining RWTH in 2011, she held a teaching position in the Department of Electrical and Computer Engineering at The University of Auckland. Her research interests are in mm-wave networking, efficient spectrum sharing paradigms, cognitive and cooperative communication, self-organizing and distributed networks, and telecommunications policy.
\end{IEEEbiography}

\begin{IEEEbiography}[{\includegraphics[width=1in,height=1.25in,clip,keepaspectratio]{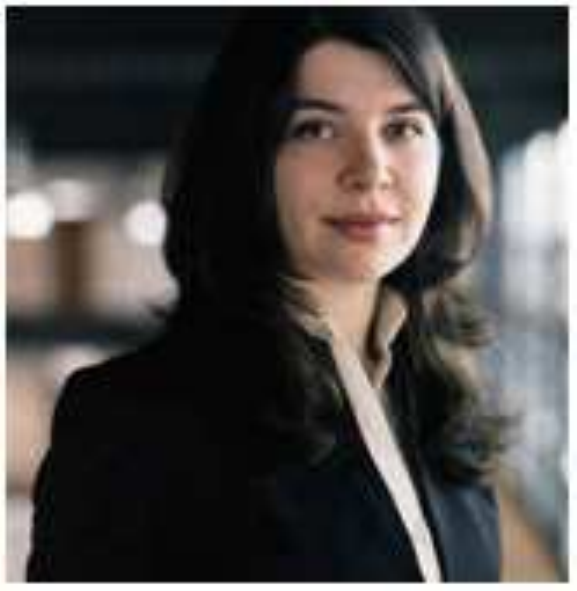}}]{Marina Petrova}
is an assistant professor and a head of the Self-Organized  Networks  research group in the Faculty of Electrical Engineering and Information Technology at RWTH Aachen University.  Her research focuses on  system-level studies  of  future  wireless  systems, modeling  and  prototyping  of  protocols  and  solutions for heterogeneous and dense wireless networks, and mm-wave communication. Dr. Petrova holds a degree in engineering and telecommunications from University Ss.  Cyril and Methodius, Skopje and a Ph.D from RWTH Aachen  University,  Germany. She was a TPC co-Chair of DySPAN 2011 and a TPC co-Chair of SRIF’14 in conjunction with SIGCOMM. She is currently serving as an editor of IEEE Wireless Communications Letters and IEEE Transactions of Mobile Computing.
\end{IEEEbiography}





\end{document}